\documentclass[prd,reprint,nofootinbib,notitlepage,aps,tightenlines,
preprintnumbers,amsmath,amssymb,amsfonts,showpacs,superscriptaddress]{revtex4-1}
\pdfoutput=1

\usepackage[hyperindex,breaklinks]{hyperref}
\usepackage{graphicx,epstopdf}
\usepackage{natbib,bm}
\usepackage[usenames,dvipsnames]{color}
\usepackage{slashed}    
\usepackage{hyperref}
\usepackage{breakurl}
\usepackage{multirow}
\usepackage{bbold}

\def\be{\begin{equation}}
\def\ee{\end{equation}}
\def\ba{\begin{eqnarray}}
\def\ea{\end{eqnarray}}
\def\ge{\mathrel{\raise.3ex\hbox{$>$\kern-.75em\lower1ex\hbox{$\sim$}}}}
\def\la{\mathrel{\raise.3ex\hbox{$<$\kern-.75em\lower1ex\hbox{$\sim$}}}}

\def\simgt{\mathrel{\raise.3ex\hbox{$>$\kern-.75em\lower1ex\hbox{$\sim$}}}}
\def\simlt{\mathrel{\raise.3ex\hbox{$<$\kern-.75em\lower1ex\hbox{$\sim$}}}}

\newcommand{\fr}[2]{\frac{#1}{#2}}

\newcommand{\nc}{\newcommand}

\nc{\gone}{\bar g_{\pi NN}^{(1)}}
\nc{\gzero}{\bar g_{\pi NN}^{(0)}}
\nc{\al}{\alpha}
\nc{\ga}{\gamma}
\nc{\de}{\delta}
\nc{\ep}{\epsilon}
\nc{\ze}{\zeta}
\nc{\et}{\eta}
\nc{\ka}{\kappa}
\nc{\rh}{\rho}
\nc{\si}{\sigma}
\nc{\ta}{\tau}
\nc{\up}{\upsilon}
\nc{\ph}{\phi}
\nc{\ch}{\chi}
\nc{\ps}{\psi}
\nc{\om}{\omega}
\nc{\Ga}{\Gamma}
\nc{\De}{\Delta}
\nc{\La}{\Lambda}
\nc{\Si}{\Sigma}
\nc{\Up}{\Upsilon}
\nc{\Ph}{\Phi}
\nc{\Ps}{\Psi}
\nc{\Om}{\Omega}
\nc{\ptl}{\partial}
\nc{\del}{\nabla}
\nc{\ov}{\overline}
\nc{\newcaption}[1]{\centerline{\parbox{15cm}{\caption{#1}}}}
\nc{\us}{U(1)$_S$}

\def\beq{\begin{equation}}
\def\eeq{\end{equation}}
\def\bmat{\begin{displaymath}}
\def\emat{\end{displaymath}}
\def\bear{\begin{eqnarray}}
\def\eear{\end{eqnarray}}
\def\ba{\begin{eqnarray}}
\def\ea{\end{eqnarray}}
\def\bery{\begin{array}}
\def\ery{\end{array}}
\def\bit{\begin{itemize}}
\def\eit{\end{itemize}}
\def\ben{\begin{enumerate}}
\def\een{\end{enumerate}}
\def\btab{\begin{tabular}}
\def\etab{\end{tabular}}
\def\btbl{\begin{table}}
\def\etbl{\end{table}}
\def\bfig{\begin{figure}[htb]}
\def\efig{\end{figure}}
\def\bpic{\begin{picture}}
\def\epic{\end{picture}}

\def\ga{\mathrel{\raise.3ex\hbox{$>$\kern-.75em\lower1ex\hbox{$\sim$}}}}
\def\la{\mathrel{\raise.3ex\hbox{$<$\kern-.75em\lower1ex\hbox{$\sim$}}}}
\def\gappeq{\mathrel{\rlap {\raise.5ex\hbox{$>$}}
{\lower.5ex\hbox{$\sim$}}}}
\def\lappeq{\mathrel{\rlap{\raise.5ex\hbox{$<$}}
{\lower.5ex\hbox{$\sim$}}}}

\def\gyr{{\rm \, G\kern-0.125em yr}}
\def\mev{{\rm \, Me\kern-0.125em V}}
\def\gev{{\rm \, Ge\kern-0.125em V}}
\def\tev{{\rm \, Te\kern-0.125em V}}

\begin{document}

\preprint{EFI-14-13}
 
\title{Leptophobic Dark Matter at Neutrino Factories}

\author{Brian Batell}
\affiliation{Enrico Fermi Institute and Department of Physics, University of Chicago, Chicago, IL 60637, USA}

\author{Patrick deNiverville}
\affiliation{Department of Physics and Astronomy, University of Victoria, 
Victoria, BC V8P 5C2, Canada}

\author{David McKeen}
\affiliation{Department of Physics, University of Washington, Seattle, WA 98195, USA}

\author{Maxim Pospelov}
\affiliation{Department of Physics and Astronomy, University of Victoria, 
Victoria, BC V8P 5C2, Canada}
\affiliation{Perimeter Institute for Theoretical Physics, Waterloo, ON N2J 2W9, 
Canada}

\author{Adam Ritz}
\affiliation{Department of Physics and Astronomy, University of Victoria, 
Victoria, BC V8P 5C2, Canada}


\begin{abstract}
\noindent High-luminosity fixed-target neutrino experiments present a new opportunity to search for light sub-GeV dark matter and associated new forces. 
We analyze the physics reach of these experiments to light leptophobic dark states coupled to the Standard Model via gauging the $U(1)_B$ baryon current. When the baryonic vector is light, and can decay to dark matter, we find that the MiniBooNE experiment in its current beam-dump configuration can extend sensitivity to the baryonic fine structure constant down to $\alpha_B\sim 10^{-6}$. This is significantly below the existing limits over much of the sub-GeV mass range currently inaccessible to direct detection experiments.
\end{abstract}
\maketitle

\section{Introduction}

A variety of gravitational phenomena strongly suggest the existence of dark matter (DM), which in its simplest form is a new stable weakly interacting elementary particle. This has motivated a broad experimental program to detect non-gravitational DM interactions, including direct searches for DM-nucleus scattering, indirect searches for DM annihilation products, and accelerator-based searches for missing energy. 
In recent years there has been a growing appreciation that fixed-target experiments provide a complementary approach to DM detection, with superior sensitivity to light sub-GeV DM interacting with ordinary matter via a light mediator particle. The potential of using high-intensity proton-beam fixed-target experiments, such as those employed to study neutrino oscillations, was highlighted and studied in Refs.~\cite{Batell:2009di,deNiverville:2011it,deNiverville:2012ij}, and a dedicated run in beam-dump mode to search for DM with the MiniBooNE experiment at the Fermi National Accelerator Laboratory is currently underway~\cite{Dharmapalan:2012xp}. More recently, the possibility of using electron-beam fixed-target experiments to search for DM has been investigated~\cite{Izaguirre:2013uxa,Diamond:2013oda,Izaguirre:2014dua}. These proposals are part of a broader effort to utilize high-intensity electron and proton fixed-target experiments, as well as high-luminosity meson 
factories, to study the physics of DM and more general hidden sectors (see, {\it e.g.} Refs.~\cite{Hewett:2012ns,Kronfeld:2013uoa,Essig:2013lka,pospelov2008,
Batell:2009yf,Essig:2009nc,Reece:2009un,Bjorken:2009mm,Freytsis:2009bh,Batell:2009jf,Freytsis:2009ct,Essig:2010xa,Essig:2010gu,McDonald:2010fe,Williams:2011qb,Abrahamyan:2011gv,Archilli:2011zc,Lees:2012ra,Davoudiasl:2012ag,Kahn:2012br,Andreas:2012mt,Essig:2013vha,Davoudiasl:2013jma,Morrissey:2014yma}).

The studies of Refs.~\cite{Batell:2009di,deNiverville:2011it,deNiverville:2012ij,Dharmapalan:2012xp,Izaguirre:2013uxa,Diamond:2013oda,Izaguirre:2014dua} have largely focused on scenarios in which DM couples to the Standard Model (SM) through a dark photon, a new massive gauge boson that kinetically mixes with the ordinary photon~\cite{Holdom:1985ag}. The DM thus primarily couples to the electromagnetic current, leading to a rich phenomenology with multiple probes involving both leptonic and hadronic systems. Such a model is well motivated on effective field theory grounds since kinetic mixing provides one of the few renormalizable ``portal"  interactions, and is viable from a phenomenological and cosmological standpoint~\cite{Pospelov:2007mp,deNiverville:2012ij}. 
Moreover, because the electromagnetic current automatically conserves many of the important symmetries ($CP$, parity, flavor), and does not couple 
to neutrinos, the resulting DM-SM interaction strength may exceed the strength of standard weak interactions without immediately running into 
strong constraints imposed by flavor physics and tests of discrete symmetries. 
However, given our ignorance regarding the structure of the DM couplings to ordinary matter, it is certainly worthwhile to explore the phenomenology of alternative models. In particular one can easily contemplate scenarios in which the mediator coupling DM to the SM is primarily hadrophilic and leptophobic, or vice-versa. Such scenarios underscore the necessity of a broad experimental program making use of both proton and electron beams. 

In this paper we investigate scenarios of sub-GeV DM in which the mediator couples dominantly to quarks, {\it i.e.} is leptophobic, and as such is uniquely suited for studies in proton fixed-target experiments. 
As we will motivate below, the specific model we consider is based on a local  $U(1)_B$ baryon number symmetry,  which, like the kinetic mixing portal, is phenomenologically safe since the corresponding current conserves all approximate symmetries of the SM.
In this model, the DM is charged under $U(1)_B$, and the baryonic gauge boson serves as the mediator coupling the DM to the SM. We provide a detailed treatment of DM production and scattering relevant for proton fixed-target experiments, and estimate the sensitivity of the ongoing beam-dump run at MiniBooNE~\cite{Dharmapalan:2012xp} to this model.
As we demonstrate, MiniBooNE will have the capability to cover significant new regions of parameter space in this model, with sensitivity to the baryonic fine structure constant at the level of $\alpha_B \sim 10^{-6}$. 

We begin in Section~\ref{sec:model} by describing a low energy effective theory containing a local $U(1)_B$ baryon number symmetry under which DM is charged. We examine several important topics, including gauge anomalies, effective couplings to hadronic states, cosmology, and existing experimental constraints. In Section~\ref{sec:fixed} we investigate the phenomenology of this model at proton fixed-target experiments. We outline the general detection strategy, describe in detail an improved DM production model, and provide a general treatment of the DM-nucleon elastic scattering. Our estimates for the sensitivities achievable with the dedicated MiniBooNE beam-dump run are presented in Section~\ref{sec:results}. Finally, our conclusions and outlook are presented in Section~\ref{sec:outlook}. Several appendices contain additional technical details.

\section{Leptophobic Dark Matter and Gauged \boldmath{$U(1)_B$}}
\label{sec:model}

We are interested in scenarios in which the interactions of light DM, $\chi$, with the SM are communicated through a new boson that dominantly couples to quarks.
Scalar bosons will generally have suppressed couplings to the lightest quark generations, implying poor detection prospects in proton-beam fixed-target experiments. Thus, in the simplest models of a scalar singlet $S$, coupled to the SM via a tri-linear Higgs portal $SH^\dagger H$, one expects the effective 
coupling of $S$ to nucleons be ${\cal O}(10^{-3} \theta)$, where $\theta$ is the mixing angle with the Higgs state. Given that one typically has constraints on $\theta$ below the $10^{-2}$ level from flavor physics, 
(see, {\it e.g.}~\cite{Batell:2009di,Batell:2009jf}), the effective coupling of $S$ to nucleons does not 
exceed $10^{-5}$, and thus is very difficult to reach directly. 

We therefore focus on a new vector boson with couplings to quarks. Without complicated model building in the flavor sector, the absence of tree level flavor changing neutral currents implies that the quark couplings should be generation independent. Furthermore, to allow renormalizable Yukawa couplings of the quarks to the SM Higgs boson, the charges of the left- and right-handed quarks should be equal. These considerations lead to a model containing a vector boson coupled to the baryon current. The most straightforward realization of such a scenario is to consider the vector boson, $V_B^\mu$, to be a fundamental gauge boson of a local $U(1)_B$ baryon number symmetry 
\cite{Nelson:1989fx,Rajpoot:1989jb,Foot:1989ts,He:1989mi,Carone:1994aa,Carone:1995pu,Bailey:1994qv,FileviezPerez:2010gw,Graesser:2011vj,FileviezPerez:2011pt,Duerr:2013dza,Dobrescu:2014fca,Tulin2014}. 

As is well known,  a model with a local $U(1)_B$ symmetry suffers from 
gauge anomalies, and therefore must be regarded as a non-renormalizable effective field theory with an ultraviolet cutoff $\Lambda_{\rm UV}$~\cite{Preskill:1990fr}. 
(Our requirement of building a leptophobic model, as motivated above, prevents us from extending this gauge symmetry to leptons 
to cancel the anomalies via, {\em e.g.} 
$U(1)_B\to U(1)_{B-L} $.)
The upper bound on $\Lambda_{\rm UV}$ can be estimated from the three loop vector boson self energy diagram and is well above the weak scale for the mass and coupling parameters explored in this study. 
At or below this scale, new states must enter to render the theory consistent at the quantum level, with the simplest possibility being a perturbative completion with new chiral fermions 
that cancel the anomalies. Such fermions may obtain large masses through Yukawa couplings to the SM Higgs boson or through couplings to the spontaneous symmetry breaking sector of $U(1)_B$. We note that a variety of constructions exist in the literature for anomaly free UV completions of a local $U(1)_B$ symmetry~
\cite{Nelson:1989fx,Rajpoot:1989jb,Foot:1989ts,He:1989mi,Carone:1994aa,Carone:1995pu,Bailey:1994qv,FileviezPerez:2010gw,Graesser:2011vj,FileviezPerez:2011pt,Duerr:2013dza,Dobrescu:2014fca}. For a given UV completion, there will inevitably be additional constraints from high energy accelerator data. Since our focus in this work is on GeV-scale phenomenology, the precise details of the UV completion will not be relevant to our discussion, and we will therefore focus on a low energy effective theory of a local $U(1)_B$ symmetry under which the DM $\chi$ is charged.

The Lagrangian of the low energy effective theory is given by
\begin{eqnarray}
\label{eq:L1}
{\cal L} & = & {\cal L}_\chi - \frac{1}{4}(V_B^{\mu\nu})^2 + \frac{1}{2}m_V^2 (V_B^\mu)^2 \nonumber \\
    && \qquad\qquad - \frac{\kappa}{2} V_B^{\mu\nu} F_{\mu\nu} +  g_B V_B^\mu J^B_\mu  + \cdots,  \\
{\cal L}_\chi & = & 
\begin{cases}
i \bar \chi \not \!\! D \chi - m_\chi \bar \chi \chi,  ~~~~~~~ ({\rm Dirac ~ fermion ~ DM})\\
|D_\mu \chi|^2 - m^2_\chi |\chi|^2,~~~~({\rm Complex ~ scalar ~ DM})
\end{cases} \nonumber 
\end{eqnarray}
where $D = \partial - i g_B q_B V_B$, with $g_B$ ($q_B$) the $U(1)_B$ gauge coupling (charge),
$J_B^\mu \equiv \tfrac{1}{3} \sum_i \bar q_i \gamma^\mu q_i$  is the baryon current (with the sum over all quark species),  and the ellipses denote terms related to the sector responsible for spontaneously breaking $U(1)_B$, the details of which will not be important for us below. 
Note that we have included a kinetic mixing term, with strength $\ka$,  in the Lagrangian~(\ref{eq:L1})~\cite{Holdom:1985ag}, which is allowed by all of the symmetries of the theory.\footnote{\label{fn:1}
Below we will present numerical results for both the $U(1)_B$ model and, for comparison, the pure vector portal model. 
The latter can be formally recovered from the Lagrangian (\ref{eq:L1}) by taking the limit $g_B\rightarrow 0$, $g'\equiv g_Bq_B\neq 0$, and $\ka\neq 0$. 
See Refs.~\cite{Batell:2009di,deNiverville:2011it,deNiverville:2012ij,Dharmapalan:2012xp} for further studies of the pure vector portal model in this context. 
} In the physical basis, the vector boson couplings to quarks are
\begin{eqnarray}
\label{Lcurrent}
{\cal L} & \supset & 
V_B^\mu \left( g_B J_\mu^B -\kappa e J_\mu^{EM} \right) , 
\end{eqnarray}
where the electromagnetic current is defined as $J_{EM}^\mu \equiv  \sum_i Q_f \bar f_i \gamma^\mu f_i$ (with the sum over all electrically charged fermions). 
Kinetic mixing can lead to a relevant deformation of the phenomenology provided $\kappa e \gtrsim g_B$. 
In models where $\kappa$ is generated radiatively, one expects to find $\kappa \sim e g_B/(16\pi^2)$.  

It is important to mention that ${\cal L}_\chi$ may contain several states coupled to the baryonic current, 
$\sum_j(i \bar \chi_j  {\not \!\! D} \chi_j - m_{\chi_j} \bar \chi_j \chi_j) $, including very light 
neutrino-like states. Models of this type were already discussed in Refs.~\cite{Pospelov:2011ha,Pospelov:2013rha,Pospelov:2012gm} 
(see also \cite{Harnik:2012ni}), where such light
states were called ``baryonic neutrinos" $\nu_b$ due to their coupling to $V_B$. Mixing with active 
neutrinos and elastic scattering on nuclei via $V_B$ exchange creates novel signatures of $\nu_b$ relevant 
for the interpretation of DM direct detection signals, provided that the interaction strength is stronger 
than the usual weak interactions. Therefore, light nearly massless dark states from the
$\chi$ sector represent an interesting physics target. As we will observe later, although they cannot 
constitute the cosmological DM, such states can be instrumental in constructing a realistic model of thermal relic DM based on $U(1)_B$. 
Moreover, the fixed-target signatures of massless $\nu_b$ states and those of very light DM  ($m_\chi \sim {\cal O}({\rm few~MeV})$) are identical, and therefore our work will also provide a method for 
constraining $\nu_b$ models. For the sake of clarity, we will denote all nearly massless states endowed with $U(1)_B$ charge as $\nu_b$, reserving the label $\chi$ for the DM. 

Since we are interested in physics below the GeV scale, it is necessary to determine the couplings of $V_B$ to mesons.  We will employ two approaches in the description of these couplings. For processes with energies below the $\rho$ meson mass $m_\rho$, we obtain the couplings of $V_B$ to the pseudoscalar mesons through the standard procedure of gauging the chiral Lagrangian. There are two distinct contributions to the $V_B$ couplings in the chiral Lagrangian. 
The first contribution arises from replacing partial derivatives of the pion Goldstone field $U$ with covariant ones, 
$\partial U \rightarrow \partial U - i V_B [ Q_B , U]$,
where the generator $Q_B = (g_B/3)\mathbb{1} -\kappa e \,Q_{EM}$, with $Q_{EM} = {\rm diag}(\tfrac{2}{3}, - \tfrac{1}{3}, -\tfrac{1}{3})$.
We observe that in the limit $\kappa \rightarrow 0$, $V_B$ does not couple to the mesons through the covariant derivative.  
The second contribution arises due to the axial anomaly and is described by the gauged Wess-Zumino-Witten (WZW) term~\cite{Wess:1971yu,Witten:1983tw} 
(see also Refs.~\cite{Kaymakcalan:1983qq,Harvey:2007ca,Hill:2008mp} for useful discussions). 
These couplings are present even in the limit $\kappa \rightarrow 0$. 
For example, the coupling of a neutral pseudoscalar meson to a photon and a vector boson $V_B$ is given by
\begin{align}
\label{Laxial}
{\cal L} &\supset -  \frac{1}{16 \pi^2 f_\pi} \epsilon_{\mu\nu\alpha\beta}F^{\mu\nu} V_B^{\alpha\beta}\Big[ 
e (g_B-\kappa e) \pi^0 \nonumber\\
& \qquad\qquad + \tfrac{1}{\sqrt{3}} e (g_B-\kappa e) \eta_8 +
2\sqrt{\tfrac{2}{3}} e (-\kappa e) \eta_0   
 \Big],
\end{align}
This coupling will mediate one of the dominant dark sector production mechanisms at proton fixed-target experiments via pseudoscalar meson decays, 
{\it e.g.}  $\pi^0\rightarrow \gamma V_B$. A detailed treatment of this subject is presented in Appendix~\ref{app:Pdecay}.

We will also be interested in production  due to vector meson mixing, for which we employ the vector meson dominance (VMD) prescription. 
Following Ref.~\cite{Kaymakcalan:1983qq}, we can write the mixing between the vector $V_B$ and the vector mesons $\rho, \omega, \phi$ as
\begin{align}
{\cal L} &\supset  \frac{\sqrt{2} }{g} V_B^\mu \Big[  (-\kappa e) \, m_\rho^2 \, \rho_\mu \, + \, \tfrac{1}{3} \, (2 g_B - \kappa e  )  \, m_\omega^2 \, \omega_\mu \nonumber\\
& \qquad\qquad\qquad - 
 \tfrac{\sqrt{2}}{3} \, (-g_B -\kappa e) \, m_\phi^2 \, \phi_\mu \Big].
\end{align}
The production of DM through vector meson mixing with $V_B$ is treated in Appendix~\ref{app:Xdecay}.

\subsection{Cosmology}

There are several challenges that the minimal model described by the Lagrangian~(\ref{eq:L1}) 
faces if one insists on $\chi$ being a viable thermal relic DM candidate. 
Besides the usual difficulty of obtaining a sufficient ($\sim$pb) 
annihilation cross section, such light DM states are 
strongly constrained by the precise measurements of the temperature anisotropies of the cosmic microwave background (CMB) radiation~\cite{Padmanabhan:2005es,Lopez-Honorez:2013cua,Galli:2013dna}. 
If the annihilation occurs into visible SM states other than neutrinos,
these constraints typically rule out a thermal relic with $s$-wave annihilation for the sub-GeV DM masses of interest in this work.
Systematically exploring the range of viable cosmologies is not crucial for this paper, and we limit our discussion to three distinct 
possibilities. 

{\em Scenario~1.} A natural model based on $U(1)_B$ achieves the correct DM abundance via annihilation to neutrino-like states, 
$\chi \bar \chi \to V_B^* \to \nu_b\bar \nu_b$, and in addition via $\chi \bar \chi \to V_B V_B \to \nu_b \bar{\nu}_b \nu_b \bar{\nu}_b$ if $m_\chi > m_{V}$.
Annihilation to these light new states completely avoids problems with energy injection during or after recombination, 
as $\nu_b$'s are not capable of ionizing Hydrogen due to their 
weak interaction with matter. In addition, it is possible to generate the required annihilation rate. For example, $\sigma v\sim$ pb 
 for  $m_{V} < m_\chi$ can be achieved by choosing $\alpha_B^2 \sim 10^{-11} (m_\chi/100 ~{\rm MeV})^2$. In the opposite 
case, $m_{V}> m_\chi$, $\alpha_B$ would need to be slightly larger and, most relevant for our discussion,  in both 
cases $g_B^2/m_V^2\equiv G_B$ would necessarily be larger than the weak Fermi constant $G_F$.

Significant sensitivity to this model comes from CMB or BBN determinations of the dark radiation energy density, 
traditionally parametrized via the effective number of neutrino degrees of freedom $N_{\rm eff}$. The naive shift of $N_{\rm eff}$                                                                                                                                                                                                                                                             in this model with $\nu_b$ is $\Delta N_{\rm eff}=1$, but the actual change might be smaller, depending on the precise time of $\nu_b$ decoupling
\cite{Pospelov:2011ha}.  
In any event, this increase to the effective number of neutrinos is not completely excluded, and furthermore this parameter 
can be additionally adjusted via new light states that decay to electrons and photons after neutrino decoupling, thereby lowering $N_{\rm eff}$.

{\em Scenario~2.}  Another minimal scenario involves scalar DM $\chi$, with annihilation aided by the ``baryonic Higgs" $h_B$, {\em i.e.} a particle 
accompanying the spontaneous breaking of $U(1)_B$. With the mass hierarchy,
 $2m_V > 2m_\chi > m_V +m_{h_B}$, it is easy to see that $\chi \chi^\dag \to 2V_B$ is kinematically forbidden, while 
the Higgs-strahlung process $\chi \chi^\dag \to V_B^*\to V_B\,h_B$ is allowed. Importantly, for scalar DM, the latter
process is necessarily $p$-wave. 
As a consequence, the CMB bounds on energy injection can be evaded due to inefficient late time annihilation. The requisite size of 
the annihilation cross section is easily achieved by an appropriate choice of $\alpha_B$. A potential problem 
for this construction is the relatively long-lived $h_B$, that would have to deplete its abundance before the start of BBN via {\em e.g.} the
$V_B h_B\to \pi^0\gamma$ co-annihilation process (see the corresponding discussion in Ref.~\cite{Pospelov:2010cw}) 
or via the two-loop decay $h_B\to 2\gamma$. It is also possible to 
achieve an accelerated decay of $h_B$ via the SM Higgs -- $U(1)_B$ Higgs portal. Because of the chosen mass hierarchy, the production of 
$\chi \chi^\dag$ in fixed-target experiments necessarily proceeds via an off-shell $V_B$.

{\em Scenario~3.}  Finally, there are always classes of models where the correct DM abundance of $\chi$ is achieved via portals which differ from 
$J^B_\mu V_B^\mu$.
One example involves a new light scalar particle $\phi$ that couples to the DM through a Yukawa interaction, ${\cal L} \supset \phi \bar \chi (a+i b \gamma^5) \chi$, with $a$, $b$, real parameters. In the regime $m_\phi < m_\chi < m_V$,  the dominant annihilation process is $\chi \bar \chi \rightarrow \phi \phi$, which will proceed in the $p$-wave if either $a$ or $b$ vanishes. The $\phi$ particle can decay to $e^+e^-$ pairs through a small Higgs portal coupling, ${\cal L }\supset A \phi H^\dag H$. We have checked, for instance, that for $m_\phi \sim 10$ MeV and $A \sim 1$ MeV,  the $\phi$ lifetime is less than one second, the effective $\phi$ coupling to electrons is consistent with Supernova cooling constraints, and the contribution to the $\phi$ mass from electroweak symmetry breaking is subdominant.

We trust that the existence of these three classes of scenarios will convince the reader that $U(1)_B$-based thermal relic DM models are possible, and we turn  next
to the existing constraints on $U(1)_B$ gauge bosons.

\subsection{Existing constraints}
\label{app:constraints}

\begin{figure*}
 \centerline{
 \includegraphics[width=0.45\textwidth]{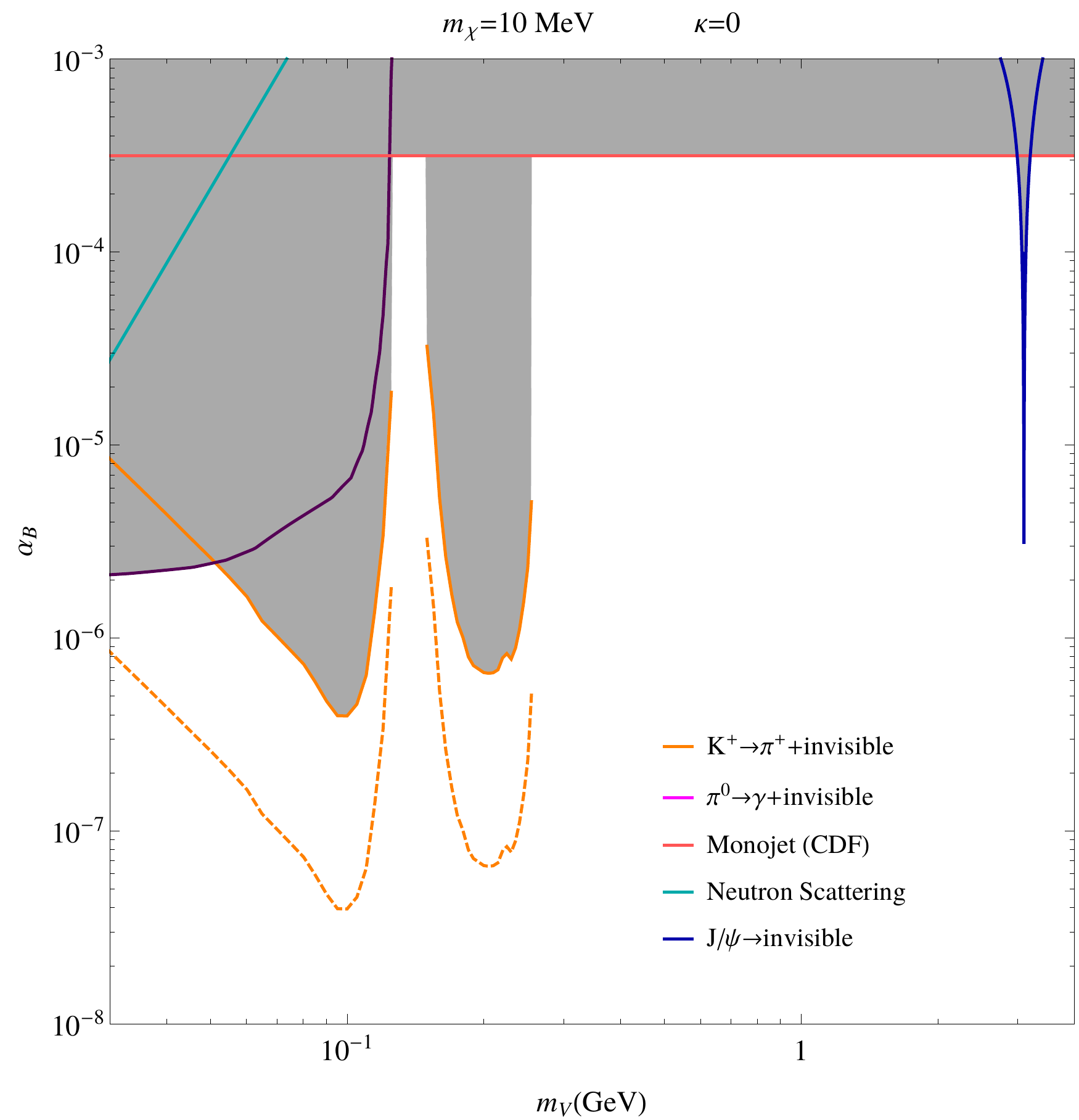}
  \hspace*{0.3cm} \includegraphics[width=0.45\textwidth]{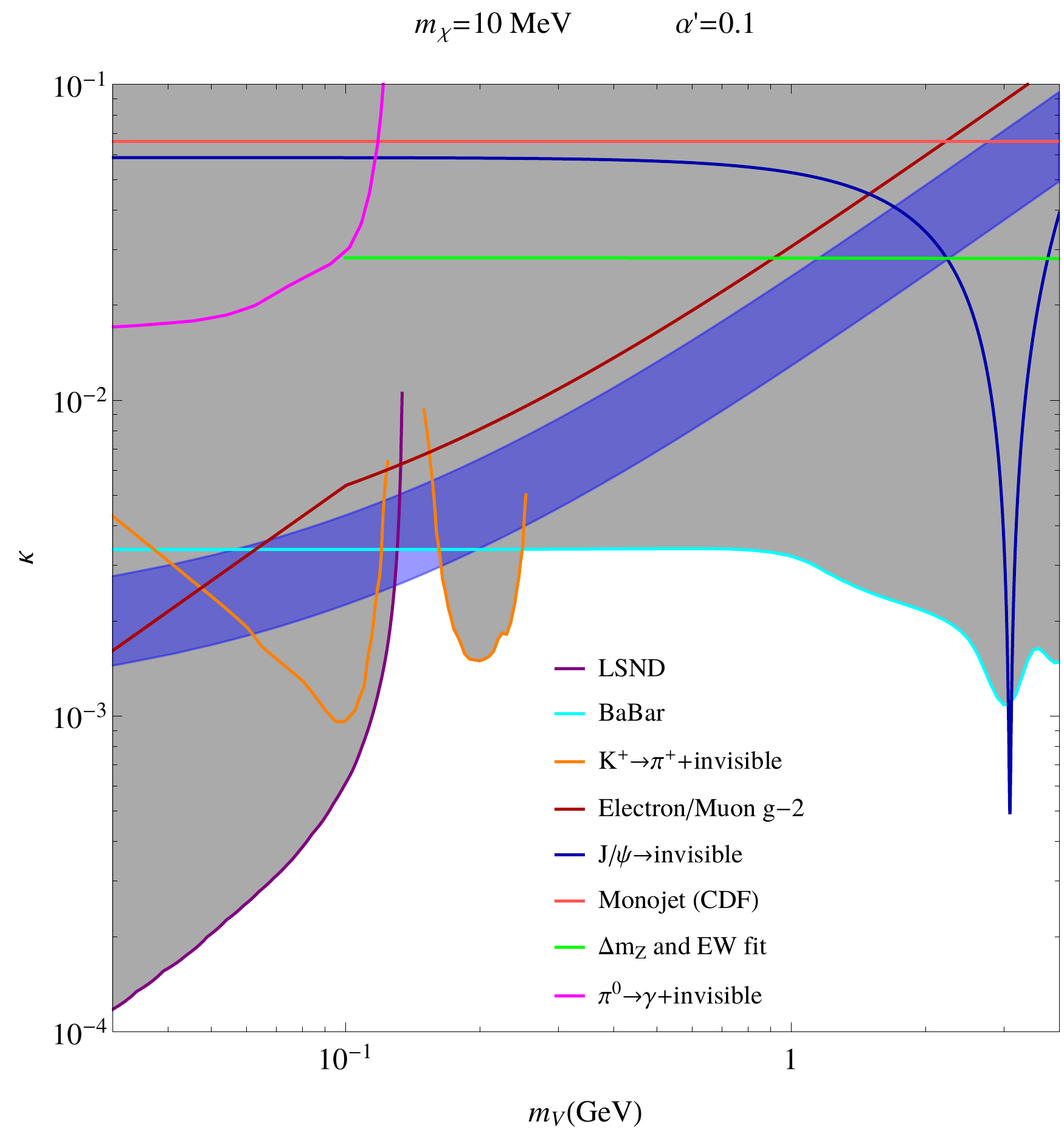}
 }
 \caption{\footnotesize 
 Existing constraints on DM model parameter space. 
 The left plot shows the constraints on the $U(1)_B$ model in the $m_V-\alpha_B$ plane for a DM mass $m_\chi = 10$ MeV and vanishing kinetic mixing $\kappa = 0$. 
 The shaded region is excluded by existing constraints. 
 The constraints shown are from limits on 
 $\pi^0 \to \gamma + {\rm invisible}$ \cite{atiya92}, 
 $J/\psi \to {\rm invisible}$ \cite{ablikim2007}, 
 $pp \to {\rm jet} + {\rm invisble}$ (labeled monojet) \cite{shoemaker2011}, 
 and neutron scattering~\cite{Barbieri1975} . 
 The limit from the $K^+\rightarrow \pi^+ \nu\bar\nu$ branching ratio measurement \cite{Kplus} is also shown under two possible assumptions on the IR cutoff: 1) $\Lambda_{IR} = 4 \pi f_\pi$ (solid orange), and 2) $\Lambda_{IR} = m_\rho$ (dashed orange). 
For comparison, the right plot shows the constraints on the pure vector portal model with $m_\chi = 10$ MeV and $\alpha'=0.1$. 
 In this model there are additional constraints originating from the sizable leptonic couplings: 
 excessive contributions to electron and muon $g{-}2$ \cite{pospelov2008,bouchendira2010,hanneke2010,aoyama2012,endo2012}, 
 a monophoton search by BaBar (labeled BaBar sensitivity) \cite{aubert2008, Izaguirre:2013uxa,Essig:2013vha}, 
 and deviations in precision electroweak measurements\cite{Hook2010}. 
 The blue band through the parameter space marks where the scenario brings theory and experiment into better than $3\sigma$ agreement  for muon $g{-}2$ \cite{pospelov2008}.
 }
 \label{fig:constraints}
\end{figure*}

In addition to the cosmological constraints, various terrestrial particle physics experiments have sufficient sensitivity to exclude portions of parameter space for the model. A number of limiting contours are shown in Fig.~\ref{fig:constraints}, and discussed below. We separate the discussion into those with specific sensitivity to $g_B$, and those which rely on kinetic mixing $\ka$ with the electromagnetic current.

\bigskip
\noindent {\bf Constraints on the baryonic coupling.} We list below the constraints with sensitivity to $g_B$ in the parameter regions of interest:

$\bullet$ {\it Rare decays with missing energy}:- Certain rare decays have significant sensitivity for both kinetic mixing and the baryonic portal. The limit on $\pi^0\rightarrow \gamma V_B$ from the Brookhaven alternating gradient synchrotron \cite{atiya92}, with the branching ratio discussed below in (\ref{Pdecay}), imposes competitive constraints at low mass for both portals. 

A stronger limit arises from the Brookhaven E949 measurement of the tiny branching fraction of $K^+\rightarrow \pi^+ \nu\bar\nu$ \cite{Kplus}, interpreted as $K^+\rightarrow \pi^+V_B$ \cite{pospelov2008}. The rate calculation needs to be generalized to account for the baryonic portal, so we include some details here. In general there are both short- and long-distance contributions. However, while the loop-induced $K^+{-}\pi^+{-}\gamma^*$ vertex can be inferred from the measured 3-body hadronic kaon decays using ChPT, since pseudosclar mesons are uncharged under $U(1)_B$, it is natural to anticipate that the long-distance contribution is suppressed. The short-distance contribution is dominated by the GIM-suppressed $V_B$-penguin with $c-$ and $u$-quark loops. Retaining just the leading logarithm, 
\begin{align}
{\cal L}_{\rm pen} &{\simeq}V_B^\mu\bar s \gamma_\mu d  \times \sin2\theta_c \fr{G_F}{\sqrt{2}} \frac{g_B}{24 \pi^2}   \log\left[\frac{m_c^2}{m_u^2({\to}\Lambda_{IR}^2)}\right]{+}{\rm h.c.},
\end{align}
where $\theta_c$ is the Cabibbo angle, and $m_u$ in the logarithm needs to be replaced with the hadronic IR cutoff $\Lambda_{IR}$, with e.g. $m_\rho \la \Lambda_{IR} \la 4\pi f_\pi$. Since we expect the long-distance contribution to be suppressed, the sensitivity to this cutoff leads to considerable uncertainty in the result.

Allowing for both the baryonic and kinetic mixing portal couplings, the amplitude takes the form,
\begin{align}
 {\cal M}_{K\rightarrow \pi V_B} &= \frac{m_V^2}{(4\pi)^2m_K^2} (k+p)^\mu \ep^V_\mu \nonumber\\
  & \qquad\quad \times(g_B W_B(m_V^2)-e\ka W_\ka(m_V^2)),
\end{align}
where $k$ and $p$ are the kaon and pion momenta, $\ep^V$ is the polarization vector of $V_B$, and $W_\ka^2(m_V^2) \sim 3\times 10^{-12}$ \cite{pospelov2008},  $W_B^2(m_V^2) \sim 4 \times 10^{-13}$ in an approximation where $m_V^2 \ll m_K^2$ and the logarithm is cut off in the infrared at the scale $\La \sim m_\rh$. This implies
\be
 {\rm Br}(K^+{\rightarrow}\,\pi^+V_B) \sim 9\times 10^{-4}\left(\frac{g_B}{3}-e\ka\right)^2 \left(\frac{m_V}{\rm 100\;MeV}\right)^2,
\ee
in agreement, up to ${\cal O}(1)$ factors, with an earlier result of Nelson and Tetradis \cite{Nelson:1989fx}. 

For larger $V_B$ masses, there are also constraints from invisible decays of $c\bar{c}$ and $b\bar{b}$ vector mesons, through mixing in analogy with the discussion below in Appendix~\ref{app:Xdecay}. We include the constraint on Br$(J/\Ps\rightarrow {\rm invisible}) < 7\times 10^{-4}$ from BES \cite{ablikim2007}.

$\bullet$ {\it CDF constraints on monojets}:- In the low ${\cal O}({\rm GeV})$ mass range, CDF provides the most stringent constraint on monojets, $p\bar{p} \rightarrow {\rm jet}+{\rm invisible}$ \cite{shoemaker2011} (see also~\cite{An:2012va}), with $g_u < 0.026$ and $g_d < 0.04$ independent of mass for $m_V < 10\,$GeV. This limit is relevant for both kinetic mixing and the baryonic portal.

$\bullet$ {\it Angular dependence in neutron scattering}:- With very low mass  vectors $V_B$ coupling via the baryonic portal, there is an additional long-range contribution to nucleon interactions, which is constrained by studies of the angular dependence in neutron scattering. For instance, from keV neutron-Pb scattering data, $\al_B < 3.4\times 10^{-11} (m_V/{\rm MeV})^4$ for $m_V > 1\,{\rm MeV}$ \cite{Barbieri1975} (as recently discussed in \cite{Tulin2014}). 

$\bullet$ {\it Direct dark matter detection}:- For DM candidates saturating the local relic density, direct detection experiments provide strong sensitivity for $m_\ch > 1$~GeV. The non-relativistic limit of $V_B$-mediated scattering allows identification of the per-nucleon cross section, $\si_N \sim 16\pi (Z/A)^2 \al\al'\ka^2\mu_{\ch,N}^2/m_V^4$ with $Z/A\sim 1/2$ for kinetic mixing and $\si_N \sim 16\pi \al_B^2\mu_{\ch,N}^2/m_V^4$ for the baryonic current. For comparison, we show the strongest low mass direct detection limits from DAMIC \cite{DAMIC}, CDMSlite \cite{CDMSLite}, SuperCDMS \cite{superCDMS}, and LUX \cite{LUX} (ordered in increasing mass). Direct detection limits on DM-electron scattering also exist~\cite{Essig:2011nj,Essig:2012yx}, although these will be subdominant for radiatively generated kinetic mixing, $\kappa \sim e g_B/16\pi^2 $. 

\bigskip
\noindent {\bf Constraints on kinetic mixing.} When kinetic mixing with hypercharge is also present via $\ka\neq 0$, several additional constraints arise due in particular to the induced leptonic couplings:

$\bullet$ {\it Loop corrections to lepton $g-2$}:- Kinetic mixing with $\ka\neq 0$ leads to a one-loop vector contribution to $g-2$ for the electron and muon. Regions for which $g-2$ deviates by more than 5$\si$ from the experimental value are excluded \cite{pospelov2008,bouchendira2010,hanneke2010,aoyama2012,endo2012}. However, for the muon, this correction can also ameliorate the disagreement between theory and experiment. The blue band in the plots indicates the parameter range for which the additional loop correction restores better than 3$\si$ agreement with the SM, and defines an interesting benchmark level of sensitivity.

$\bullet$ {\it Elastic scattering at LSND}:- An important limit on kinetic mixing at low mass arises from an analysis of the LSND measurement of elastic neutrino scattering on electrons \cite{Auerbach:2001wg}. A limit was placed on non-standard scattering contributions which, with the large $\sim 10^{23}$ POT dataset and production via neutral pion decay to DM through an on-shell vector mediator, allows a strong constraint to be placed on this light DM scenario as discussed in more detail in \cite{deNiverville:2011it}. 

$\bullet$ {\it BaBar Monophotons}:- For kinetic mixing, one of the most significant constraints comes from the BaBar monophoton search, which can be interpreted in terms of invisibly decaying vectors which are produced in association with a single photon in $e^+e^-$ collisions \cite{aubert2008,Izaguirre:2013uxa,Essig:2013vha}. This relies crucially on the single photon trigger, and allows sensitivity over the full mass range.

$\bullet$ {\it $\De m_Z$ and EW fit}:- Kinetic mixing with hypercharge also has an impact on the $\gamma-Z$ alignment after electroweak symmetry breaking. The ensuing shift of $m_Z$, along with the precision of the global electroweak fit, also imposes a significant (and essentially mass-independent) limit \cite{Hook2010}.

$\bullet$ {\it Rare visible decays}:- Visible decays of the vector provide relatively weak limits with kinetic mixing (and even weaker limits for the baryonic portal). For the kinetic mixing parameters studied here, the vector decays promptly and the limits imposed by dark photon searches at MAMI, BaBar, APEX and KLOE \cite{Hewett:2012ns,KLOE-2,MAMI-2} are suppressed by the visible branching fraction to leptons, $\ka^2\al/\al'$, as the dominant decays are invisible (to $\ch\chi^\dagger$). For the baryonic portal, the dominant visible meson decays are even further suppressed, either due to the need for decays to three-body final states ({\it e.g.} 3 pions), and/or through anomaly-mediated channels (as recently discussed in \cite{Tulin2014}). The limits in each case are subleading to the other constraints shown in Fig.~\ref{fig:constraints}.

\section{Signatures at fixed-target neutrino experiments}
\label{sec:fixed}

We now investigate the sensitivity of proton fixed-target experiments to the model of leptophobic DM described in the previous section and by the Lagrangian (\ref{eq:L1}). In this section we will outline the basic detection strategy that can be employed by neutrino experiments such as MiniBooNE. We will provide a detailed overview of the production of relativistic DM in the primary proton-target collisions, as well as a treatment of the DM-nucleon scattering. Our estimates for the sensitivity of the dedicated MiniBooNE beam-dump run to leptophobic DM will be presented in Section~\ref{sec:results}. We note that much of the discussion here follows that of our earlier works of Refs.~\cite{Batell:2009di,deNiverville:2011it,deNiverville:2012ij}, although we will present a more comprehensive treatment of the DM production model, which will extend the reach of MiniBooNE to DM masses closer to 1 GeV.

\subsection{Detection Strategy}

In neutrino experiments such as MiniBooNE, an intense proton beam is directed onto a fixed target, resulting in strong production of hadrons. An extended decay volume downstream of the target allows the pions to decay in flight, resulting in a large flux of neutrinos. 
Being weakly interacting, neutrinos can travel unimpeded through the dirt, potentially oscillating along the way, and then scatter in the detector via charged and neutral current processes.

If DM $\chi$ couples to quarks, as happens in the leptophobic model (\ref{eq:L1}) considered here, then both the mediator and the DM can be copiously produced in the primary proton-target collisions. There are a number of DM production mechanisms, including the decays of secondary mesons $\pi^0, \eta, \eta'$, mixing of the vector mediator $V_B$ with vector-mesons $\rho, \omega, \phi$, and through direct perturbative QCD production. This results in a relativistic flux of DM directed along the beam line. Just like neutrinos, DM interacts very weakly with ordinary matter, and can thus reach the near detector and scatter elastically with nucleons through a $t$-channel $V_B$ exchange. Thus, the signature of DM at these experiments is a neutral current nucleon scattering event. 

Since the signature is neutral current-like scattering, neutrinos constitute a significant background to the DM signal. There are several strategies that can be employed to combat this beam-related background~\cite{Dharmapalan:2012xp}.
The kinematic differences in the nucleon recoil energy and angular spectrum can be exploited through a dedicated analysis. This requires a detailed understanding of the neutrino background spectrum and will not be pursued further here. Secondly, one can utilize precise timing information to search for the scattering events that are out of time with the proton beam spill, which would be expected for heavier DM particles, $m_\chi \gtrsim 100$ MeV, which have a delayed arrival at the detector relative to the neutrinos. 
Finally, one can dramatically reduce the neutrino flux by directing the protons onto a beam dump, with no decay volume. In this case the charged mesons are absorbed or stopped before they decay, resulting in a smaller and more isotropic neutrino flux, while the DM production mechanisms are unaltered. MiniBooNE is currently carrying out a beam-dump run, with the expectation of reducing the neutrino flux by a factor  of $\sim50$.

\subsection{Dark Matter Production}

\begin{figure*}
\centerline{\includegraphics[width=0.45\textwidth]{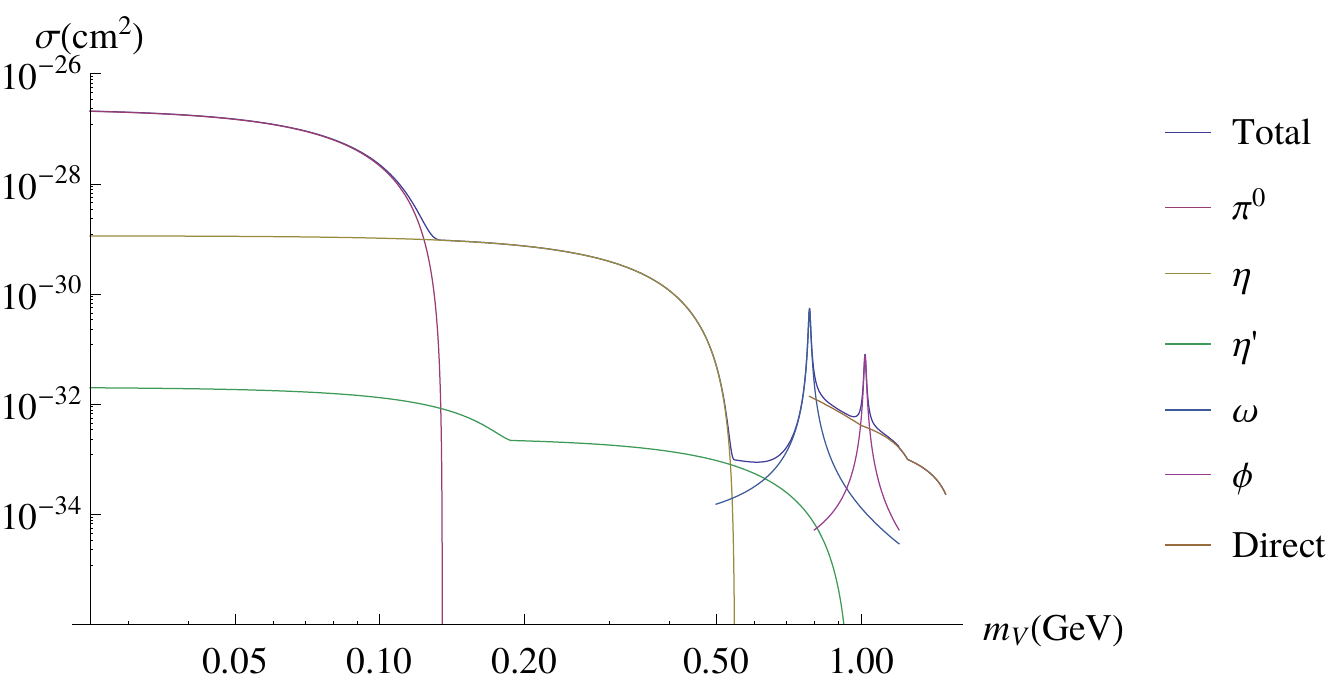}\hspace{0.2in}\includegraphics[width=0.45\textwidth]{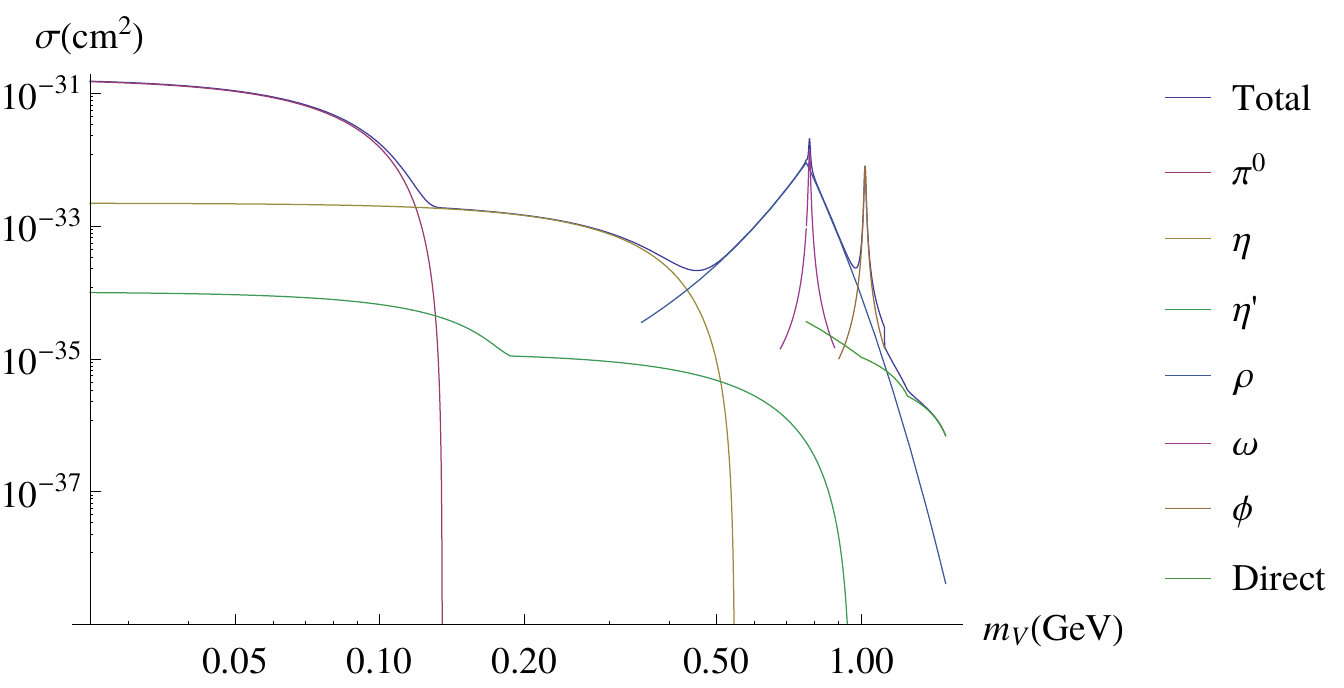}}
 \caption{\footnotesize Vector boson production cross sections $\sigma(p p \rightarrow \chi \chi^\dag + \dots)$ as a function of $m_V$ 
for MiniBooNE energies, broken into individual production channels. We have fixed the DM mass to $m_\chi= 10$ MeV. We show the results for the $U(1)_B$ model with $\alpha_B = 10^{-4}$, $\kappa = 0$ (left), and the pure kinetic mixing portal model with $\ka = 10^{-3}$ (right) for comparison. The blue line represents the sum of all production channels considered.}
 \label{fig:production}
\end{figure*}

We now turn to a quantitative treatment of the DM production model. We will specialize to the case of production at MiniBooNE, although the description can easily be modified where appropriate for other experiments. In general, we would like to determine as precisely as possible $\si(pp(n)\rightarrow \ch\ch^\dagger + \cdots)$, or equivalently $\si(pp(n)\rightarrow V_B^*+\cdots)$ since Br$(V_B\rightarrow \chi\ch^\dagger)\approx 1$ in all cases studied here. Since $V_B$ is a narrow resonance for the parameters of interest, if we denote $q^2$ as the invariant mass of $V_B^*$, then the cross section is well-approximated by $q^2\sim m_V^2$. At MiniBooNE, the low $4.2\,$ GeV $pp(n)$ center-of-mass energy requires us to consider multiple hadronic production modes for $m_V < 1$\,GeV. 

We will focus on three classes of production processes: 1) secondary meson decay, 2) vector meson mixing, and 3) direct QCD production for sufficiently large $q^2\sim m_V^2$. Although beyond the scope of this work, one can also contemplate production of DM through bremsstrahlung-like radiation of the vector mediator from the proton beam, and it would be worthwhile to investigate this mechanism in the future. 
\begin{table}[t]
\begin{tabular}{|c||c|c|c|c|c|}
\hline
 {\rm Meson} & $\eta/\pi^0$ & $\eta'/\pi^0$ & $\rho/\pi^0$ & $\omega/\pi^0$ & $\phi/\pi^0$ \\ \hline
$\frac{\sigma}{\sigma_{\pi^0}} = \frac{\si}{90\,{\rm mb}}$ & 1/30 &  1/300 & 0.05 & 0.046 & 1/150  \\
\hline
\end{tabular}
\caption{\footnotesize 
Estimates of the production cross sections for the 8.9 GeV beam at MiniBooNE \cite{Teis:1996kx,compilation}. 
The number of particles produced is given by $N\sim{\rm POT}\times \si L_{nuc}  n_N$, where $L_{nuc}\sim$~1 interaction length, 
and $n_N$ the number density. The numbers quoted are for the beryllium target, but can be rescaled for the iron absorber.}
\label{table1}
\end{table}

$\bullet$ {\it Secondary meson decay}:- 
For low mass vectors, the dominant production mode is via radiative decay of pseudoscalar mesons $\varphi=\pi^0,\eta,\eta'$ \cite{Batell:2009di,deNiverville:2011it}. We take $\si(pp(n)\rightarrow V_B + \cdots)\sim \si(pp(n)\rightarrow \varphi+\cdots)\times {\rm Br}(\varphi \rightarrow \gamma V_B)$, and 
\begin{equation}
\label{eq:ratio-P}
\frac{{\rm Br}(\varphi \rightarrow \gamma V_B)}{{\rm Br}(\varphi \rightarrow \gamma\gamma)} =
 2 \left(c_\varphi \frac{g_B}{e} - \kappa \right)^2  \left( 1 - \frac{m_V^2}{m_{\varphi}^2}\right)^3,
\end{equation}
where $c_\varphi \approx \{1,0.61,-0.12\}$ for $\varphi = \pi^0, \eta, \eta'$. Further details of the computation are presented in Appendix~\ref{app:Pdecay}. 
Estimated production rates for the pseudoscalar mesons at MiniBooNE are summarized in Table~\ref{table1}.

$\bullet$ {\it Vector meson mixing}:- For $m_V$ close to the mass of a vector meson $X=\rho,\omega,\phi$, resonant production via mixing can be important \cite{Morrissey:2014yma}. 
In principle, this requires an off-shell treatment of both $X$ and $V_B$, to account for the full spectral shape. 
However, there is little ({\it e.g.} Drell-Yan) data available for the relevant kinematic range, and we will focus on one tractable contribution that corresponds to taking $\si(pp(n) \rightarrow V_B^*+\cdots)\sim \si(pp(n) \rightarrow X+\cdots)\times {\rm Br}(X\rightarrow V_B^*\rightarrow \ch\ch^\dagger)$. This relation can be derived in the narrow-width approximation for the vector meson resonance, and one can compute the branching ratio
\begin{widetext}
\begin{eqnarray}
\frac{{\rm Br}(X \rightarrow \chi \bar \chi)}{{\rm Br}(X\rightarrow e\bar e) } \! &= & \! r_\chi \left( c_X \frac{g_B}{e}-\kappa  \right)^2 \left(\frac{g_B q_B}{e} \right)^2 \frac{m_X^4}{(m_X^2 - m_V^2)^2 +m_V^2 \Gamma_V^2} \left(1+ a_\chi\frac{m_\chi^2}{m_X^2} \right) \left(1-\frac{4 m_\chi^2}{m_X^2} \right)^{1/2},
\label{eq:Xdecay}
 \end{eqnarray}
 \end{widetext}
where $c_{X} = \{0,2,-1\}$  for $X= \{\rho,\omega, \phi\}$, while $r_\ch = 1$, $a_\chi = 2$ (Dirac fermion $\ch$), or $r_\ch=1/4$, $a_\chi = -4$ (scalar $\ch$). 
In practice, the $X$ width is usually much larger than the $V_B$ width, so to better approximate the spectral shape we will broaden the effective resonance width, $\Gamma_V\rightarrow \Gamma_{\rm eff}\sim \Gamma_X$. (In the case of $\rho$, we also modify the spectral shape as a Breit-Wigner distribution does not provide a good fit to higher energy Drell-Yan data.) Further calculational details are presented in Appendix \ref{app:Xdecay}. 
Estimated production rates for the vector mesons are again summarized in Table~\ref{table1}.

$\bullet$ {\it Direct QCD production}:- For $m_V$ above roughly a GeV, we use direct parton-level production via $q\bar{q}\rightarrow V_B$, and work with the narrow width approximation for $V_B$,
\begin{widetext}
\begin{align}
\sigma (pp(n) \rightarrow V_B) &= \frac{\pi}{3m_V^2} \sum_q \left(\frac{g_B}{3} - \kappa e Q_q\right)^2  \int_{\tau}^1 \frac{dx}{x} \tau \left[ f_{q/p}(x)f_{\bar{q}/p(n)}\left(\frac{\tau}{x}\right)+f_{\bar{q}/p}(x)f_{q/p(n)}\left(\frac{\tau}{x}\right)\right],
\end{align}
\end{widetext}
where $\ta=m_V^2/s$. We use the CTEQ6.6 parton distribution functions $f_{q/p(n)}(x)$ and $f_{\bar{q}/p(n)}$ setting the scale $Q=m_V$. The uncertainties for $m_V\sim 1\,$GeV at MiniBooNE energies are likely ${\cal O}(1)$, but we find that the rates are not that large in practice so higher-order corrections are not likely to significantly modify the conclusions. Further details, including the full differential distributions, are discussed in \cite{deNiverville:2012ij}.

In Fig.~\ref{fig:production} we display the $V_B$ production cross sections for the various channels described above for the $U(1)_B$ model with $\kappa = 0$, as well as the pure vector portal model for comparison (see the Footnote~\ref{fn:1} for an explanation).

\begin{table*}
\begin{tabular}{|c|c|c|c|c|c|c|c|}
\hline
  POT & $E_{\rm beam}$ & $L$  & $A_{\rm det}$ & $L_{\rm det}$ & $n_{CH_2}$ & Fiducial mass & $\ep_{\rm eff}$ \\ \hline
  $2 \times10^{20}$ & 8.9~GeV &  541m & $1.2\times10^6\,{\rm cm}^2$ & 11.5m & $9 \times10^{23}\,{\rm cm}^{-3}$ &  $~450$~tons & 0.35 \\
\hline
\end{tabular}
\caption{\footnotesize A summary of the relevant MiniBooNE parameters used in this work; see the text for further details and notation.}
\label{table2}
\end{table*}

\subsection{Dark Matter-Nucleon Elastic Scattering}

Once produced in the primary collisions, the DM can be detected through its elastic scattering signature in the near detector. 
The DM-nucleon differential elastic scattering cross section can be written as
\begin{widetext}
\begin{equation}
\frac{d\sigma_{\chi N \rightarrow \chi N}}{d E_\chi} =  \alpha_B \, q_B^2 \,  \frac{  \tilde F^2_{1,N} A(E,E_\chi) + \tilde F^2_{2,N} B(E,E_\chi) +  \tilde F_{1,N}  \tilde F_{2,N} C(E,E_\chi)   }{(E^2 - m_\chi^2)(m_V^2 + 2 m_N (E- E_\chi))^2},
\label{E:scat}
\end{equation}
\end{widetext}
where $E$($E_\chi$) is the incoming (outgoing) dark matter energy, $m_N$ is the nucleon mass, $Q^2=2m_N(E-E_\ch)$ is the momentum transfer, the expressions for the form factors  $\tilde F_{(1,2),N}$ are given in Eq.~(\ref{eq:FF}), and the kinematic functions $A,B,C$ depend on the DM spin and are given in Eqs.~(\ref{eq:ABC-scalar}) and (\ref{eq:ABC-fermion}) for a complex scalar  and Dirac fermion, respectively. Further details of the scattering computation are presented in Appendix~\ref{app:scatter}.

\section{Results}
\label{sec:results}

To generate estimates of the signal rate, the next step is to simulate DM production distributions, so that the specific geometric and energy cuts relevant for MiniBooNE can be incorporated.

\subsection{Production and Scattering Simulation}
\label{ssec:production_sim}

The momentum and angular distributions of the parent mesons were simulated by sampling the MiniBooNE Sanford-Wang meson production fits \cite{AguilarArevalo:2008yp} using an acceptance-rejection method. The $\pi^0$ distribution was approximated using the mean of MiniBooNE's $\pi^+$ and $\pi^-$ fits, a procedure which, according to previous studies (see e.g \cite{amaldi1979,jaeger1974}), produces a fit in reasonably good agreement with the measured $\pi^0$ distribution. For the other mesons considered, we instead use MiniBooNE's $K^0_s$ fit in order to obtain some estimate of how the momentum distribution changes for a particle of much higher mass than that of a pion, though in practice the two distributions are quite similar. The pseudoscalar mesons thus produced are decayed into vectors $V_B$ and other final state particles, while the vector mesons are replaced with $V_B$ particles of the same momentum and angle. The vectors $V_B$ are decayed into $\chi \chi^\dag$ pairs, providing a set of DM trajectories emanating from the MiniBooNE target. Direct production is handled in a similar manner, but it samples $V_B$ and $\chi$ decay angles from the production distribution detailed previously in \cite{deNiverville:2012ij}. Dark matter particles possessing trajectories that intersect the MiniBooNE detector are recorded for later use in calculating the MiniBooNE event rate. This procedure is performed for all relevant production channels for a given $V_B$ mass.

The set of DM trajectories produced for each production channel $A$ are summed over in order to calculate the DM event rate in the MiniBooNE detector. For mesons, we use
\begin{widetext}
\begin{align}
\label{eq:eventsum}
N_{\chi N\to\chi N,A} &= \ep_{\rm eff} N_A {\rm Br}(A\to V_B +\cdots) {\rm Br}(V_B \to \chi \chi^\dag) n_{{\rm CH}_2} 
 \times \frac{1}{J_A} \sum_j l_j \sigma_{\chi {\rm CH_2} \to \chi {\rm CH_2}}(E_j),
\end{align}
\end{widetext}
where $\ep_{\rm eff}$ is the detection efficiency, $N_A$ is the number of mesons $A$ produced in the MiniBooNE target for a given POT, $n_{CH_2}$ is the number density of mineral oil in the MiniBooNE detector, $J$ is the total number of DM trajectories generated for production channel $A$, $l_j$ is the length of intersection of the DM trajectory $j$ and the MiniBooNE detector, and $\sigma_{\chi {\rm CH_2} \to \chi {\rm CH_2}}(E_j)$ is defined as
\begin{widetext}
\begin{equation}
\label{eq:}
\sigma_{\chi {\rm CH_2} \to \chi {\rm CH_2}} = \int_{0.1 {\rm GeV}^2}^{1.6 {\rm GeV}^2} dQ^2 \left( 6 C_{\nu p,C} \frac{d\sigma_{\chi p \to \chi p}}{dQ^2} + 6 C_{\nu p,C}\frac{d\sigma_{\chi n \to \chi n}}{dQ^2} + 2 C_{\nu p,H}\frac{d\sigma_{\chi p \to \chi p}}{dQ^2} \right),
\end{equation}
\end{widetext}
where the $Q^2$ dependent efficiencies $C_{\nu (n,p),(C,H)}$ are as listed in Appendix B.2 of \cite{AguilarArevalo:2010cx}. For direct production, we make the substitution $N_A {\rm Br}(A\to V_B+\cdots) \to N_{V_B}$. The estimate of the total event rate is calculated by adding the results of the individual production channels together.

\subsection{Sensitivity}

\begin{figure*}
 \centerline{\includegraphics[width=0.45\textwidth]{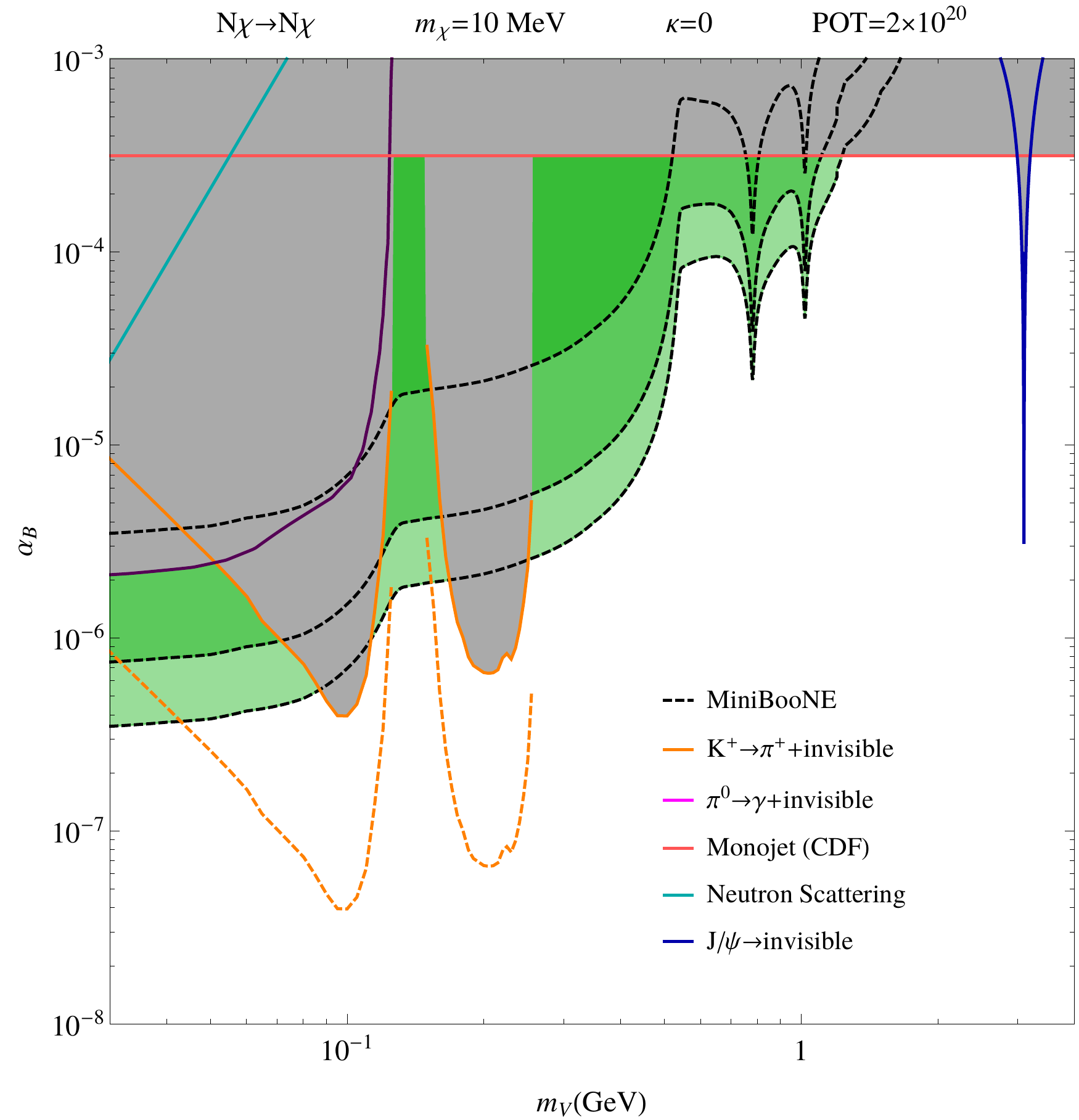} \hspace*{0.3cm} \includegraphics[width=0.45\textwidth]{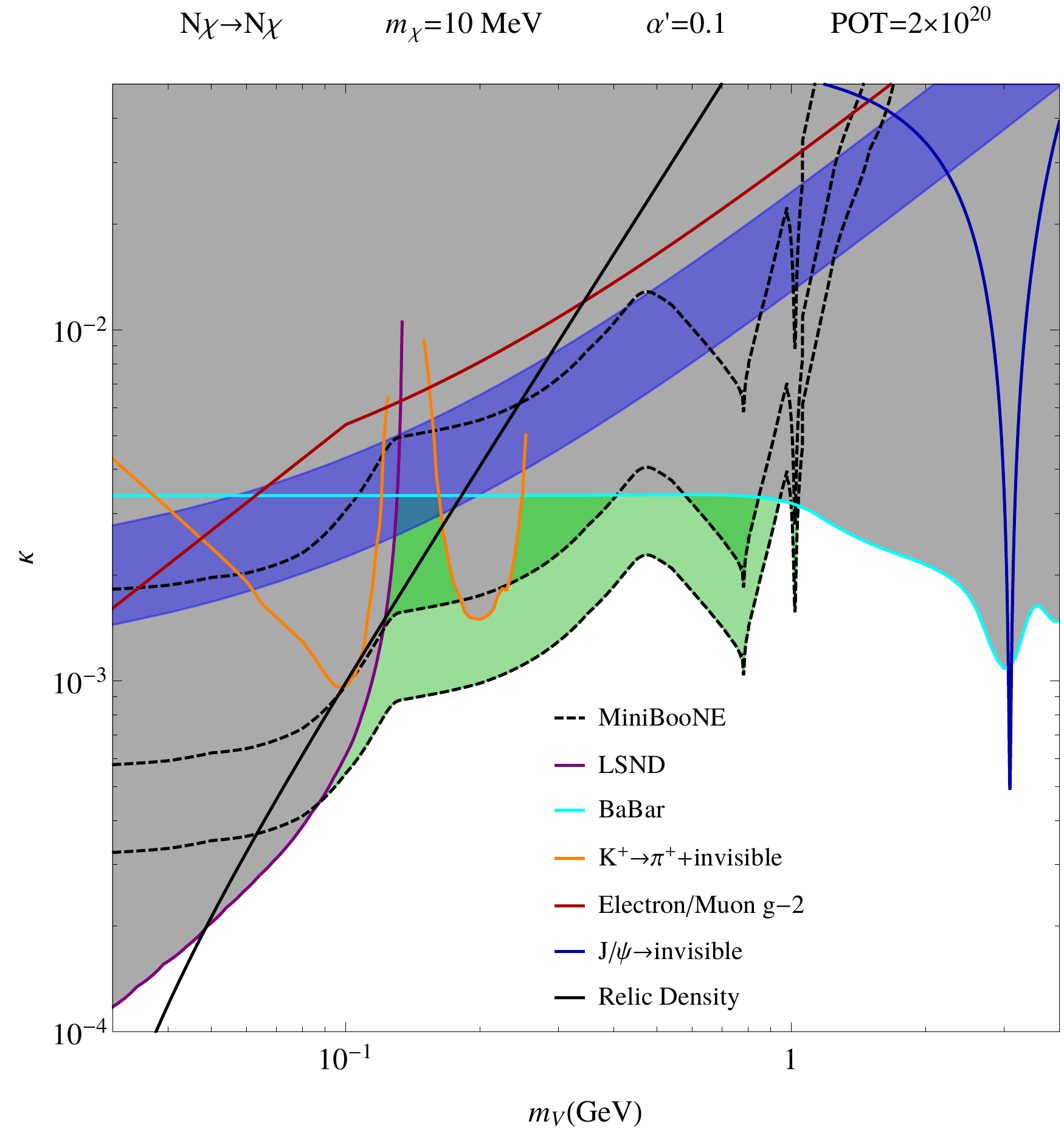}}
 \caption{\footnotesize 
 Sensitivity contours for the MiniBooNE beam-dump run (green), with the three contour regions corresponding to 1 event (light), 10 events (medium) and 1000 events (dark). In grey are exclusions from other sources, which are detailed in Section~\ref{app:constraints}. The left panel displays the sensitivity for the $U(1)_B$ model in the $m_V-\alpha_B$ plane, assuming a DM mass of $m_\chi = 10$ MeV and vanishing kinetic mixing, $\kappa = 0$. For comparison, the right panel  displays the sensitivity for the pure vector portal model for $m_\chi = 10$ MeV and $\alpha'=0.1$. The black line through the parameter space (labeled Relic density) traces the combination of parameters that reproduce the observed matter density of the universe. \cite{deNiverville:2012ij}. }
 \label{fig:minimx10mev}
\end{figure*}

The parameters relevant for MiniBooNE in its current beam-dump run configuration are shown in Table~\ref{table2}, including the expected final POT to be achieved by the end of summer 2014. The efficiencies are adopted from the published neutral current analysis. With these parameters, the simulation described above was used to determine the expected number of events, and the contours are shown in a series of plots overlaid on top of the existing constraints. As described in \cite{Dharmapalan:2012xp}, use of various techniques to reduce the neutrino background should allow sensitivity to DM scattering at the 100-event level.

In the left panel of Fig.~\ref{fig:minimx10mev} we display the sensitivity of MiniBooNE to the $U(1)_B$ model in the $m_V-\alpha_B$ plane, assuming a DM mass of $m_\chi = 10$ MeV and vanishing kinetic mixing, $\kappa = 0$. The shaded green regions correspond to 1 (light), 10 (medium) and 1000 (dark)  expected DM-nucleon scattering events during the beam-dump run. We observe that MiniBooNE will be able to test a substantial region of unexplored parameter space, probing couplings as low as $\alpha_B \sim 10^{-6}$ and $V_B$ masses up to $m_V \sim 1$ GeV.

For comparison, in the right panel of Fig.~\ref{fig:minimx10mev} we display the sensitivity of MiniBooNE to the pure vector portal model for the same DM mass and $\alpha'=0.1$ (see Footnote~\ref{fn:1} for an explanation). The existing constraints from LSND, BaBar, and $K\rightarrow \pi \nu \bar \nu$ cover much of the parameter space to which MiniBooNE is sensitive. As discussed in Section \ref{app:constraints}, these constraints are essentially a consequence of the larger leptonic couplings present in the model.  However, MiniBooNE is capable of probing an interesting range of unconstrained parameters, $\kappa \sim {\rm 2} \times 10^{-3}$ and $m_{\pi^0} < m_V \lesssim  1$ GeV.

\begin{figure*}
 \centerline{\includegraphics[width=0.45\textwidth]{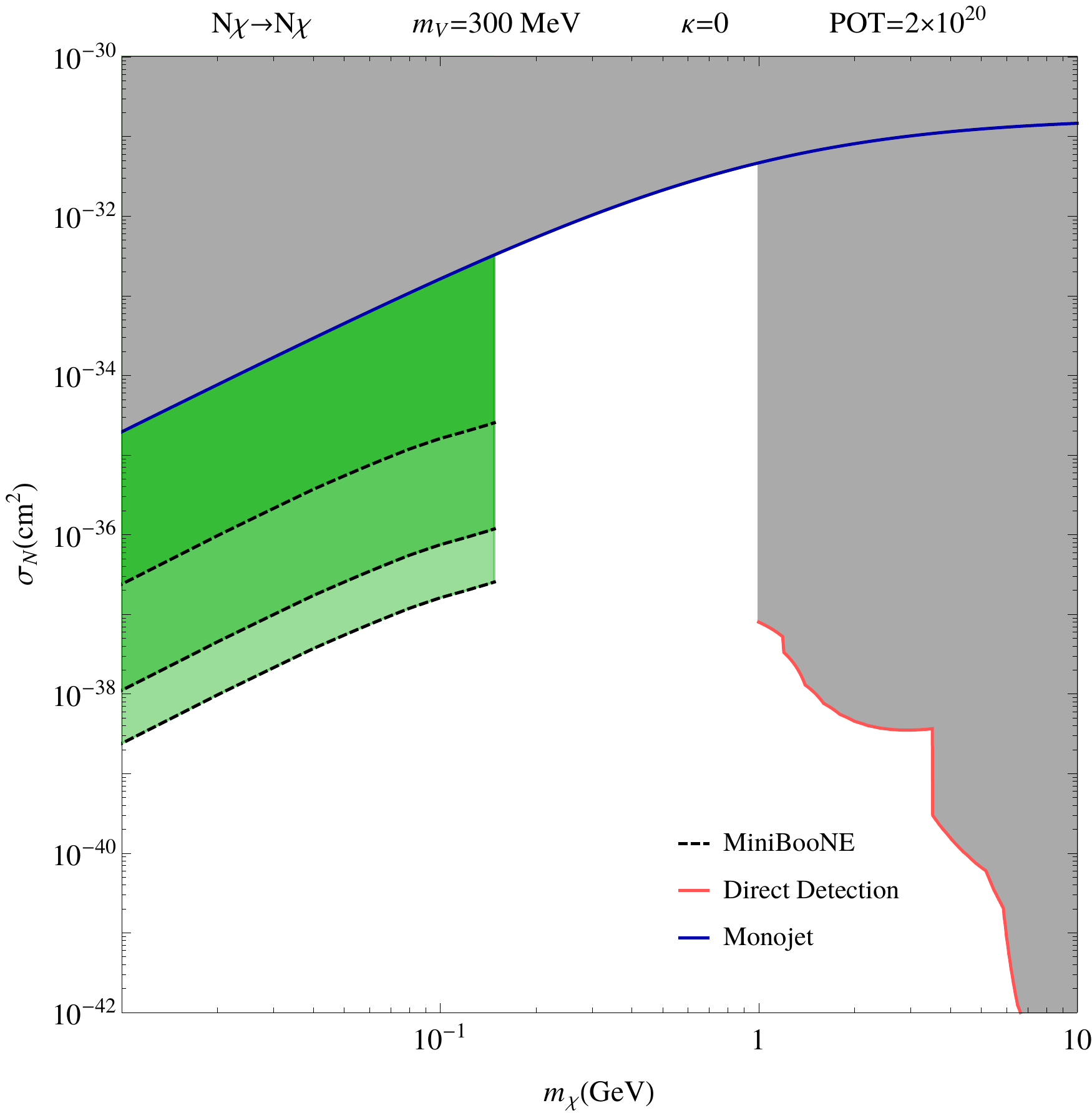} \hspace*{0.3cm} \includegraphics[width=0.45\textwidth]{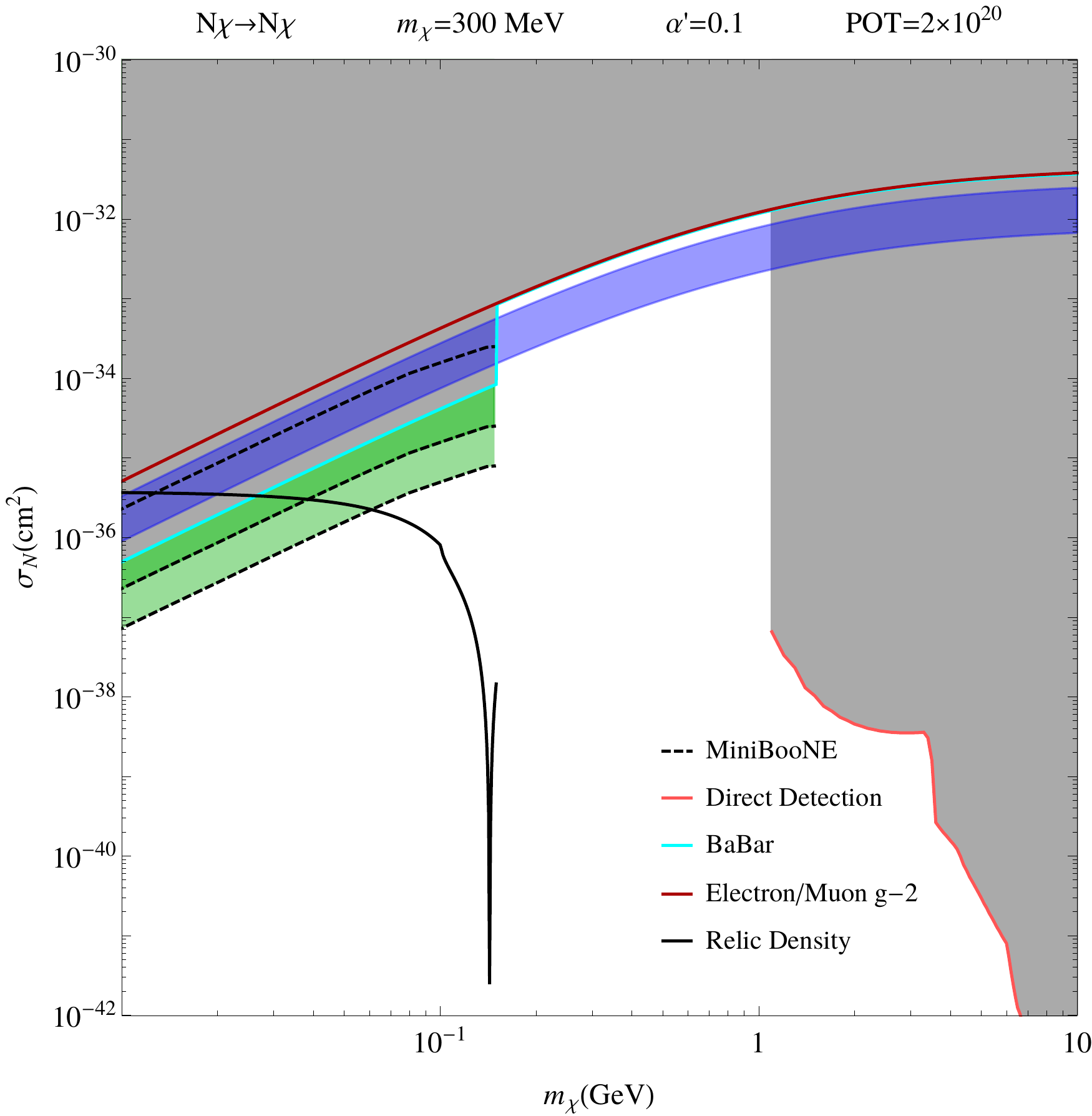}}
 \caption{\footnotesize 
 Sensitivity contours for the MiniBooNE beam-dump run (green) in the direct detection plane (see the text for a description), with the three contour regions corresponding to 1 event (light), 10 events (medium) and 1000 events (dark). In grey are exclusions from other sources, which are detailed in Section~\ref{app:constraints}. The left panel shows the sensitivity for the $U(1)_B$ model, with $m_V = 300$ MeV and vanishing kinetic mixing, $\kappa = 0$. For comparison, the right panel shows the sensitivity for the pure vector portal model, with $m_V = 300$ MeV and $\alpha' = 0.1$, with the solid black line again showing the parameters required to reproduce the observed dark matter density \cite{deNiverville:2012ij}.
}
 \label{fig:minimv300mev}
\end{figure*}

We also show in Fig.~\ref{fig:minimv300mev} the MiniBooNE sensitivities in the direct detection plane (effective spin-independent DM-nucleon cross section vs. DM mass -- see the discussion in Section \ref{app:constraints} for details on this conversion). The left panel shows the sensitivity for the $U(1)_B$ model, with $m_V = 300$ MeV and vanishing kinetic mixing, $\kappa = 0$, while the right panel shows for comparison the sensitivity for the pure vector portal model, with $m_V = 300$ MeV and $\alpha' = 0.1$. These plots highlight both the impressive capability of MiniBooNE and, more generally, the unique potential of proton-beam fixed-target experiments to probe light leptophobic DM.

Finally, let us comment on the case of sizable kinetic mixing, $\kappa e \sim g_B$, in the $U(1)_B$ model. In this case, as $\kappa$ is increased, the leptonic couplings become larger, and the constraints from LSND and BaBar, among others, become relevant. However, the DM production and scattering rates are not dramatically altered, since they primarily occur through couplings of the vector mediator to quarks.

\section{Outlook}
\label{sec:outlook}

This paper has highlighted the unique sensitivity of fixed-target neutrino experiments to leptophobic light DM scenarios. 
We focussed on a generic model in which the DM candidate interacts predominantly via coupling to the gauged baryon current. 
We have demonstrated that the MiniBooNE beam-dump run will be able to test an impressive range of model parameters
that are currently unconstrained.

Below, we remark on several important directions for further study:

\begin{itemize}
\item {\it Higher proton beam energies}: While we have focused on MiniBooNE, which uses the 9 GeV Fermilab Booster as its proton source, a number of existing and planned neutrino experiments employ higher energy proton beams. Examples include MINOS and NOvA (120 GeV protons from the Fermilab Main Injector), T2K (30 GeV protons from the JPARC synchrotron), and the CNGS facility at CERN (400 GeV protons from the CERN SPS). Looking to the future, there is the LBNE experiment, which will use an intense proton source based at Fermilab, and the SHIP program, which will use the CERN SPS beam. Future searches for light sub-GeV DM provide an important new physics motivation for these experiments, and therefore it will be critical to study the sensitivity of these facilities to the light leptophobic DM scenarios considered here. On the experimental side, we encourage the collaborations to begin to develop dedicated analyses aimed at detecting anomalous neutral current events, which could be induced by light DM states. Due to the higher proton beam energy, direct QCD production of vectors and DM will become more relevant, and heavier dark sector states of the order of 1-10 GeV can be produced. The sensitivity of these experiments to the pure vector portal model was considered previously in Ref.~\cite{deNiverville:2012ij}. It would also be useful to expand the investigation of the scattering signatures to the deeply inelastic regime, since the characteristic energies of DM particles intersecting the detector are in the tens of GeV range.  
\item {\it Visible decays of $V_B$}: 
One crucial parameter for the model is the relation between the DM mass $m_\chi$ and the vector mass $m_V$, or 
more generally the question about the existence of states lighter than $m_V/2$ charged under 
$U(1)_B$. 
In the preceding sections we have implicitly discussed both, since some of the DM production 
mechanisms did not require $2 m_\chi < m_V$.
It is easy to see that for $2 m_\chi < m_V$ the rate of visible decays ($V_B\to e \bar e $, 
$V_B \to \pi^0\gamma$, etc.) are diluted by the dominant $V_B\to \chi\bar\chi$ decay mode, while in the opposite case
$V_B \to  {\rm SM}$ proceeds unimpeded and indeed may provide a sensitive probe of the model.
The latter case also requires the absence of extra light states $\nu_b$. 
The phenomenology of the visible decays of a GeV-scale $U(1)_B$ gauge boson were recently discussed in Refs.~\cite{Dobrescu:2014fca,Tulin2014}.
 
For $2 m_\chi > m_V$ it is conceivable that the vector state coupled to the baryon 
current can be more efficiently probed directly through observation of the visible final states in its decay.  
In addition to the searches via rare meson decays discussed in Ref.~\cite{Dobrescu:2014fca,Tulin2014}, 
one can utilize proton and electron fixed-target experiments to search for $V_B$ visible decays. 
In the case of proton beams, one can employ the same $V_B$ production channels described in this paper. For a certain range of parameters, 
the vector boson will be metastable and reach the near detector before decaying. 
For vector masses below $m_{\pi^0}$, $V_B$ will decay  to either an $e^+ e^-$ pair through kinetic mixing, or to a three photon final state through an off-shell $\pi^0$, leading to distinctive electromagnetic signatures that can be searched for with MiniBooNE as well as liquid argon-based detectors such as MicroBooNE. Above the pion threshold, the vector will dominantly decay to a $\pi^0 \gamma$ state, again leading to a three photon signature. 
 
The visible decays of $V_B$ also provide an excellent physics case for electron fixed-target searches, 
and in particular the HPS experiment \cite{Stepanyan:2013tma} at Jefferson Lab. 
In this experiment, significant sensitivity should be possible via a search for displaced vertices in decays 
to electron-positron pairs. In the pure vector portal model, the experimental sensitivity extends down to $\kappa \sim 10^{-5}$ 
in the mass range $20~{\rm MeV} \la m_V \la 200~{\rm MeV} $. It is easy to see that this is precisely the range of $\kappa$ expected in the models 
with gauged $U(1)_B$, due to radiatively induced kinetic mixing 
$\kappa_{\rm ind} \sim g_B e /16 \pi^2$. Despite some uncertainty due to initial value of $\kappa$, for $\alpha_B \sim 10^{-6}$ one anticipates
$\kappa \sim 10^{-5}-10^{-4}$, right in the middle of the parameter space accessible via the displaced vertex search 
by HPS. Therefore, a significant fraction of these models could result in both the DM scattering signature and visible signatures in electron machines. 

Finally, it is also possible that the sensitivity to $m_V$ could be extended above 200 MeV via the search for visible decay modes. 
This will depend crucially on the direct production rate of $V_B$ in electron-target collisions
via the conversion of an off-shell photon, $p+\gamma^* \to p + V_B$.
Notice that this process does not require $\kappa \neq 0$ and can be induced by the baryonic current. 
Evaluating this electro-production 
mechanism and the ensuing sensitivity to the $U(1)_B$ parameter space goes beyond the scope of this paper, 
but is important as it could compete with the $\eta$ decay channels suggested in \cite{Tulin2014}.
\item {\it UV completions of local $U(1)_B$}: As discussed in Section~\ref{sec:model}, the model considered here suffers from gauge anomalies, and requires a UV completion. While a number of explicit UV complete models of gauge $U(1)_B$ exist in the literature~\cite{Nelson:1989fx,Rajpoot:1989jb,Foot:1989ts,He:1989mi,Carone:1994aa,Carone:1995pu,Bailey:1994qv,FileviezPerez:2010gw,Graesser:2011vj,FileviezPerez:2011pt,Duerr:2013dza}, it would be worthwhile to revisit this issue in the context of the light  sub-GeV DM scenario considered in this work. There will in general be additional model-dependent constraints from high energy colliders on the new heavy states responsible for the quantum consistency of the theory. For some recent discussion of this matter, see Ref~\cite{Dobrescu:2014fca}. 
\item {\it Astrophysical sensitivity}: The DM models studied here generally exhibit either suppressed or hidden annihilation channels in the late universe, in order to satisfy, for example, the constraints from the CMB. The astrophysical signatures are therefore quite limited. However, for vector mass scales at or below the supernova core temperature, $m_V \sim 30$~MeV, coupling through kinetic mixing or the baryonic portal may allow the production of DM within the core, e.g. via $NN\rightarrow NN V \rightarrow NN \bar{\ch}\ch$. For sufficiently weak coupling, e.g. $10^{-9} \lesssim \ka \lesssim10^{-6}$ \cite{Dreiner} and $10^{-19} \lesssim \al_B \lesssim 10^{-13}$ depending on the vector mass, cooling of the core via free streaming of $\ch$ is inconsistent with the observed neutrino emission from SN1987A \cite{Raffelt}. This process could be considered in more detail, but the constraints are not relevant for the larger values of $\ka$ and $\al_B$ considered here, for which DM thermalizes and only diffuses slowly from the core. Dark matter would instead form a thermal sphere and be radiated to form a diffuse SN background in the same manner as neutrinos, albeit at a much lower rate. 
\end{itemize}

\section*{Acknowledgements}

We would like to thank Andrey Afanasev, Ranjan Dharmapalan, Bogdan Dobrescu, Rouven Essig, Claudia Frugiuele, Liping Gan, 
Ashot Gasparian, Richard Hill, Zarko Pavlovic, Ze'ev Surujon, and Richard Van de Water for helpful discussions. 
The work of B.B. is supported by the NSF under Grant PHY-0756966 and the DOE under Grant DE-SC0003930.  
The work of  P.dN., M.P. and A.R. is supported in part by NSERC, Canada, and research at the Perimeter Institute 
is supported in part by the Government of Canada through NSERC and by the Province of Ontario through MEDT. 
D.M. is supported in part by the DOE under Grant No. DE-FG02-96ER40956.

\appendix
\numberwithin{equation}{section}

\section{Dark sector production}

\subsection{Pseudoscalar meson decays}
\label{app:Pdecay}

Here we compute the decays of pseudoscalar mesons to vectors, $\varphi \rightarrow \gamma V_B$, where $\varphi=\pi^0, \eta, \eta'$. 
Given the generic interactions, 
\begin{eqnarray}
{\cal L} & = & g_{\varphi \gamma\gamma} \, \epsilon_{\mu\nu\alpha\beta} \, \varphi \, \partial^\mu A^\nu \, \partial^\alpha A^\beta +  
g_{\varphi\gamma V} \, \epsilon_{\mu\nu\alpha\beta} \, \varphi\, \partial^\mu A^\nu \, \partial^\alpha V_B^\beta, \nonumber \\
&&
\end{eqnarray}
one obtains the branching ratio for $\varphi \rightarrow \gamma V_B$:
\begin{eqnarray}
\label{ratio}
\frac{{\rm Br}(\varphi \rightarrow \gamma V_B)}{ {\rm Br}(\varphi \rightarrow \gamma \gamma)  } & = & \frac{1}{2} \frac{g_{\varphi\gamma V}^2}{g_{\varphi\gamma\gamma}^2} \left( 1 - \frac{m_V^2}{m_\varphi^2}\right)^3.
\end{eqnarray}
It remains to determine the couplings $g_{\varphi\gamma\gamma}$, $g_{\varphi\gamma V}$, which arise from the gauged WZW Lagrangian.

First consider the two-photon couplings. In the $U(3)_f$ symmetric limit, the Lagrangian is given by
\begin{equation}
{\cal L} = \frac{\alpha}{2 \pi f_\pi} \epsilon_{\mu\nu\alpha\beta} \, \partial^\mu A^\nu \, \partial^\alpha A^\beta \left(  c_\pi \pi^0 + c_8 \eta_8 + c_0 \eta_0  \right),
\end{equation}
where $U(3)_f$ symmetry dictates that  $(c_\pi, c_8, c_0)=(1, \tfrac{1}{\sqrt{3}} , \sqrt{\tfrac{8}{3}})$. Fixing $f_\pi = 92.2$ MeV gives the correct prediction for the $\pi^0\rightarrow \gamma\gamma$ decay width. To reproduce the correct partial widths for the two photon decays of $\eta$ and $\eta'$, we must include two additional effects: 1)  $U(3)_f$ breaking in the form of distinct decay constants for each meson,  {\it i.e.} $f_\pi, f_8, f_0$, and 2) $\eta-\eta'$ mixing, such that the flavor eigenstates are related to the mass eigenstates by 
\begin{equation}
\eta_8  = \cos \theta \, \eta +\sin\theta \, \eta', ~~~~~~
\eta_0  = -\sin \theta \, \eta +\cos\theta \, \eta'.~
\end{equation}
We adopt the following values from Ref.~\cite{Cao:1999fs}:
\begin{equation}
\theta = -14.5^\circ,  ~~~~   f_8/f_\pi = 0.94,  ~~~~ f_0/f_8 = 1.17.
\end{equation}
The two-photon couplings are then given by
\begin{eqnarray}
g_{\pi\gamma\gamma} & = &  \frac{\alpha}{2\pi f_\pi}, \nonumber  \\
g_{\eta\gamma\gamma} & = &  \frac{\alpha}{2\pi} \left(  \frac{c_8}{f_8}\cos\theta - \frac{c_0}{f_0}\sin\theta   \right),  \nonumber \\
g_{\eta'\gamma\gamma} & = &  \frac{\alpha}{2\pi}  \left(  \frac{c_8}{f_8}\sin\theta + \frac{c_0}{f_0}\cos\theta   \right). 
   \label{eq:PGaGa-coupling}
\end{eqnarray}

Next we consider the $\varphi\gamma V_B$ couplings. Again, we start from the $U(3)_f$ symmetric terms from the WZW Lagrangian:
\begin{align}
\label{Laxial}
{\cal L} &\supset   \frac{1}{4 \pi^2 f_\pi} \epsilon_{\mu\nu\alpha\beta}  \partial^\mu A^\nu  \partial^\alpha V^\beta \Big[  
c_\pi e (g_B-\kappa e) \pi^0 \nonumber\\
& \qquad\qquad + 
c_8 e (g_B-\kappa e) \eta_8 +
c_0 e (-\kappa e) \eta_0 
 \Big].
\end{align}
Including the effects of $U(3)_f$ breaking in the decay constants and $\eta -\eta'$ mixing, we derive the couplings 
\begin{eqnarray}
g_{\pi\gamma V} &=&  \frac{\alpha}{\pi f_\pi} \left(\frac{g_B}{e} - \kappa\right),   \\
g_{\eta\gamma V} &=&  \frac{\alpha}{\pi} \left[  \frac{c_8}{f_8}\cos\theta\, \frac{g_B}{e}  - \left( \frac{c_8}{f_8}\cos\theta   - \frac{c_0}{f_0}\sin\theta  \right) \kappa
   \right],  \nonumber \\
g_{\eta'\gamma V} &=&  \frac{\alpha}{\pi}  \left[  \frac{c_8}{f_8}\sin\theta\, \frac{g_B}{e}  - \left( \frac{c_8}{f_8}\sin\theta   + \frac{c_0}{f_0}\cos\theta  \right) \kappa
   \right]. \nonumber \\
   && \nonumber
   \label{eq:PGaV-coupling}
\end{eqnarray}
Plugging Eqs.~(\ref{eq:PGaGa-coupling},\ref{eq:PGaV-coupling}) into Eq.~(\ref{ratio}), we obtain the result of Eq.~(\ref{eq:ratio-P}) in the text,
\begin{equation}
\frac{{\rm Br}(\varphi \rightarrow \gamma V_B)}{ {\rm Br}(\varphi \rightarrow \gamma \gamma)  } = 2 \left( c_\varphi \frac{g_B}{e} - \kappa  \right)^2 \left( 1 - \frac{m_V^2}{m_\varphi^2}\right)^3,
\label{Pdecay}
\end{equation}
where
\begin{eqnarray}
c_\pi & = & 1, \nonumber \\
c_{\eta} & = & \left(1-\frac{c_0}{c_8} \frac{f_8}{f_0} \tan\theta  \right)^{-1} \approx 0.61, \nonumber \\
c_{\eta'} & = & \left(1+\frac{c_0}{c_8} \frac{f_8}{f_0} \cot\theta  \right)^{-1} \approx -0.12 .
\end{eqnarray}


\subsection{Vector meson decay}
\label{app:Xdecay}
Here we compute the decays of vector mesons $X = \rho, \omega, \phi$ to DM pairs, $X\rightarrow \chi \bar \chi$. These decays occur due to $X-V_B$ mixing under the VMD hypothesis. It will be convenient to normalize the branching ratios to ${\rm Br}(X\rightarrow e^+ e^-)$, which occurs due to $X-\gamma$ mixing. Consider first the generic couplings 
\begin{equation}
{\cal L} \supset g_{XY}  X_\mu Y^\mu + g_F Y_\mu \bar F\gamma^\mu F + i g_S  Y_\mu S^*   \overleftrightarrow {\partial^\mu}   S, 
\label{eq:L-XV}
\end{equation}
where $X$ is a vector meson, $Y$ is either the photon or baryonic vector $V_B$, $F$ is the electron or Dirac fermion DM, and $S$ is the complex scalar DM. 
The partial decay widths for $X \rightarrow  \bar F  F$ and $X\rightarrow S^*  S$ are given by 
\begin{widetext}
\begin{eqnarray}
\Gamma(X\rightarrow \bar F F) & = & \frac{g_F^2 g_{XY}^2 m_X}{12 \pi} \frac{1}{(m_X^2-m_Y^2)^2 + m_Y^2 \Gamma_Y^2} \left( 1+\frac{2 m_F^2}{m_X^2} \right) 
\left( 1- \frac{4 m_F^2}{m_X^2}  \right)^{1/2},  \nonumber \\
\Gamma(X\rightarrow S^* S) & = & \frac{g_S^2 g_{XY}^2 m_X}{48 \pi} \frac{1}{(m_X^2-m_Y^2)^2 + m_Y^2 \Gamma_Y^2} 
\left( 1- \frac{4 m_S^2}{m_X^2}  \right)^{3/2}.
\end{eqnarray}
\end{widetext}

Let us now specialize to the gauged $U(1)_B$ model of interest. The $X$-photon  mixing Lagrangian is 
\begin{equation}
{\cal L} \supset  \frac{\sqrt{2} e}{g} A^\mu \left(  m_\rho^2 \rho_\mu  + \tfrac{1}{3} m_\omega^2 \omega_\mu  - \tfrac{\sqrt{2}}{3} m_\phi^2 \phi_\mu \right).
\end{equation}
We therefore identify the photon-$X$ mixing parameter $g_{XA}$, defined via Eq.~(\ref{eq:L-XV}), as 
$g_{XA} = \tfrac{\sqrt{2}e}{g}m_X^2 a_X$, 
$a_X = (1,\tfrac{1}{3},-\tfrac{\sqrt{2}}{3})$.
The $X-V_B$ mixing Lagrangian is
\begin{align}
{\cal L} &\supset  \frac{\sqrt{2} }{g} V^\mu_B \Big[  (-\kappa e) \, m_\rho^2 \, \rho_\mu \, + \, \tfrac{1}{3} \, (2 g_B - \kappa e  )  \, m_\omega^2 \, \omega_\mu \nonumber\\
 & \qquad\qquad  - 
 \tfrac{\sqrt{2}}{3} \, (-g_B -\kappa e) \, m_\phi^2 \, \phi_\mu \Big].
\end{align}
The $X-V_B$ mixing parameter $g_{XV}$ is thus
$g_{XV} = \frac{\sqrt{2}}{g} m_X^2 a_X ( c_X g_B -\kappa e)$,
where
 $a_X = (1,\tfrac{1}{3},-\tfrac{\sqrt{2}}{3})$, 
 $c_X = (0,2,-1)$.
 
We then obtain the branching ratio for 
$X\rightarrow \chi \bar \chi$ given in Eq.~(\ref{eq:Xdecay}) in the main text:
 \begin{widetext}
\begin{eqnarray}
\frac{{\rm Br}(X \rightarrow \chi \bar \chi)}{{\rm Br}(X\rightarrow e\bar e) } \! &= & \! r_\chi \left( c_X \frac{g_B}{e}-\kappa  \right)^2 \left(\frac{g_B q_B}{e} \right)^2 \frac{m_X^4}{(m_X^2 - m_V^2)^2 +m_V^2 \Gamma_V^2} \left(1+ a_\chi\frac{m_\chi^2}{m_X^2} \right) \left(1-\frac{4 m_\chi^2}{m_X^2} \right)^{1/2},
 \end{eqnarray}
 \end{widetext}
where 
$r_\chi = (1,\tfrac{1}{4})$, $a_\chi = (2,-4)$ for a Dirac fermion  and complex scalar DM. We have taken $g_F = g_S = g_B q_B$ for the DM coupling to $V_B$. 

\section{DM-nucleon scattering}
\label{app:scatter}
The computation of the DM-nucleon scattering cross sections follows the analogous computation for neutrino-nucleus scattering (see {\it e.g.}, \cite{Leitner-thesis})
 and utilizes the hypothesis of partial conservation of the axial current (PCAC). We consider the process
$\chi(p) + N(k) \rightarrow \chi(p')+ N(k')$,
 where $N = p,n$, and begin by writing the quark vector currents 
in terms of $U(3)_f$ flavor currents:
\begin{eqnarray}
J^\mu_3  &= &\bar q \gamma^\mu \frac{\lambda^3}{2} q = \frac{\bar u \gamma^\mu u - \bar d \gamma^\mu d}{2}, \nonumber \\
J^\mu_0 & = & \bar q \gamma^\mu \frac{\lambda^8}{2\sqrt{3}} q = 
\frac{\bar u \gamma^\mu u + \bar d \gamma^\mu d - 2 \bar s \gamma^\mu s}{6} ,\nonumber \\
J^\mu_{\cal S} & = &  
\bar q \gamma^\mu \left( -\frac{\lambda^8}{2\sqrt{3}} + \frac{\lambda^9}{2\sqrt{6}}  \right)q  
=\frac{\bar s \gamma^\mu  s }{2},
\label{su3current}
\end{eqnarray}
where $q^T = (u,d,s)$ and $\lambda^a$ are Gell-Mann matrices.
The couplings of $V_B$ to the light quarks are
\begin{eqnarray}
\label{Lcurrent-scatter}
{\cal L} & \supset &
V_B^\mu ( g_u \bar u \gamma_\mu u +g_d \bar d \gamma_\mu d +g_d \bar s \gamma_\mu s )
 \nonumber \\
& = &V_B^\mu \left[   (g_u{-}g_d) J_\mu^3{+}3 (g_u{+}g_d)  J_\mu^0{+}2(g_u{+}2 g_d) J^{{\cal S}}_\mu \right], \qquad
\end{eqnarray}
where we have defined 
$g_u = g_B/3 -2\kappa e/3$, $g_d = g_B/3 +\kappa e/3$.
The matrix elements of these currents between external nucleon states are:
\begin{align}
\langle k'  |  J^\mu_3 (0)  | k  \rangle&{=}\bar U(k') \left[\gamma^\mu F_1^{(v)}{+}\frac{i \sigma^{\mu\nu} q_\nu}{2 m_N} F_2^{(v)}  \right] \frac{\sigma^3}{2} U(k), \nonumber  \\
\langle k'  |  J^\mu_0 (0)  | k  \rangle  &{=}\bar U(k') \left[  \gamma^\mu F_1^{(s)}{+}\frac{i \sigma^{\mu\nu} q_\nu}{2 m_N} F_2^{(s)}  \right] \frac{\mathbf{1}}{2} U(k), \nonumber  \\
\langle k'  |  J^\mu_{\cal S} (0)  | k  \rangle  &{=}\bar U(k') \left[  \gamma^\mu F_1^{\cal S}{+}\frac{i \sigma^{\mu\nu} q_\nu}{2 m_N} F_2^{\cal S}  \right] \frac{\mathbf{1}}{2} U(k), 
 \label{U3matrix}
\end{align}
where the nucleon spinor are $U^T = (u_p, u_n)$. 
The form factors in Eq.~(\ref{U3matrix}) are functions of the momentum transfer $Q^2 = -q^2$, with $q = k' - k$.
The isovector and isoscalar form factors $F_{1,2}^{(v,s)}$ are related to the Dirac and Pauli form factors $F_{1,2}^{p,n}$ via
\begin{equation}
F_{1,2}^{(v,s)} = F_{1,2}^p \mp F_{1,2}^n, \label{DPFF}
\end{equation}
which are in turn related to the Sachs form factors,
\begin{eqnarray}
F_{1}^{p,n} & = & \frac{G_E^{p,n} + \tau \, G_M^{p,n}  }{1+ \tau }, \nonumber  \\
F_{2}^{p,n} & = & \frac{G_M^{p,n} - G_E^{p,n}  }{1+\tau }, \label{sachs1} \\
&& \nonumber 
\end{eqnarray}
where $\tau \equiv Q^2/4 m_N^2$.
The Sachs form factors are parameterized as:
\begin{eqnarray}
G_E^p(Q^2) &  =  &  G_D(Q^2), \nonumber   \\
G_E^n(Q^2) & = &  0, \nonumber \\
G_M^p(Q^2) &  =  &  \mu_p G_D(Q^2), \,~~~~~~  \mu_p = 2.793,   \nonumber \\
G_M^n(Q^2) & = &    \mu_n G_D(Q^2), ~~~~~~ \mu_n = -1.913,   \nonumber \\
G_D(Q^2) & = & \frac{1}{(1+ \frac{Q^2}{M^2})^2}, \,~~~~~ M = 0.843\, {\rm GeV}.  \label{sachs2}
\end{eqnarray}
The experimental determinations of the strange form factors
are consistent with being equal to zero~\cite{AguilarArevalo:2010cx}. 
Using the expressions above, we can compute the matrix element of the current given in (\ref{Lcurrent-scatter}) between external nucleon states,
\begin{widetext}
\begin{equation}
 \langle k' | J^\mu(0) | k \rangle = \bar u_N(k') \left[  \gamma^\mu \tilde F_{1,N} + \frac{i \sigma^{\mu\nu}  q_\nu }{2 m_N}\tilde F_{2,N}  \right] u_N(k),
\end{equation}
where $N = p, n$, and the form factors are given by 
\begin{eqnarray}
\tilde F_{(1,2),p} & = & \frac{1}{2} (g_u-g_d) F_{(1,2)}^{(v)} +  \frac{3}{2} (g_u+g_d) F_{(1,2)}^{(s)} +
 (g_u+2g_d) F_{(1,2)}^{\cal S},  \\
\tilde F_{(1,2),n} & = &  -\frac{1}{2} (g_u-g_d) F_{(1,2)}^{(v)} +  \frac{3}{2} (g_u+g_d) F_{(1,2)}^{(s)} +
 (g_u+2g_d) F_{(1,2)}^{\cal S}  .
 \label{eq:FF}
\end{eqnarray}
With these ingredients, we compute the differential cross section for DM-nucleon elastic scattering:
\begin{equation}
\frac{d\sigma_{\chi N\rightarrow \chi N}}{d E_\chi} =  \alpha_B \, q_B^2 \, \frac{  \tilde F^2_{1,N} A(E,E_\chi) + \tilde F^2_{2,N} B(E,E_\chi) +  \tilde F_{1,N}  \tilde F_{2,N} C(E,E_\chi)   }{(E^2 - m_\chi^2)(m_V^2 + 2 m_N (E- E_\chi))^2}.
\end{equation}
\end{widetext}
The functions $A,B,C$ depend on the spin of the DM. For the case of a complex scalar DM, we obtain
\begin{eqnarray}
A_{(s)} & = &  2 m_N E E_\chi - m_\chi^2 (E - E_\chi), \nonumber \\
B_{(s)} & = &  \frac{1}{4}(E - E_\chi) [  (E+ E_\chi)^2 - 2 m_N (E - E_\chi) - 4 m_\chi^2 ],   \nonumber \\
C_{(s)} & = & -(E - E_\chi) (m_N (E - E_\chi)   +  2 m_\chi^2 ),
\label{eq:ABC-scalar}
\end{eqnarray}
while for a Dirac fermion DM, we obtain
\begin{eqnarray}
A_{(f)} & = &  m_N [E (E-m_N) + E_\chi (E_\chi + m_N)  ]  - m_\chi^2 (E - E_\chi),   \nonumber \\
B_{(f)} & = & \frac{1}{2} (E- E_\chi) [2 E E_\chi +  m_N (E - E_\chi) - 2 m_\chi^2 ],  \nonumber \\
C_{(f)} & = & 2 (E - E_\chi) (m_N (E - E_\chi)   -  m_\chi^2 ) . 
\label{eq:ABC-fermion}
\end{eqnarray}

\bibliography{bpdma}

\end{document}